\documentclass[letterpaper]{emulateapj}
\usepackage{subfigure}
\newcommand{\dens}{$n_{\rm e}$}
\newcommand{\temp}{$T_{\rm e}$}
\newcommand{\oii}{[\ion{O}{2}]}
\newcommand{\sii}{[\ion{S}{2}]}
\newcommand{\cliii}{[\ion{Cl}{3}]}
\newcommand{\ariv}{[\ion{Ar}{4}]}
\newcommand{\nii}{[\ion{N}{2}]}
\newcommand{\oiii}{[\ion{O}{3}]}

\newcommand{\cmtres}{$\rm{cm}^{-3}$}

\shortauthors{Delgado-Inglada and Rodr\'iguez}
\shorttitle{C/O abundance ratios, iron depletions, and dust features in Galactic PNe}

\begin{document}

\title{C/O abundance ratios, iron depletions, and infrared dust features in Galactic Planetary 
Nebulae} 

\author{Gloria Delgado-Inglada\altaffilmark{1, 2} and M\'onica Rodr\'iguez\altaffilmark{1}}
\altaffiltext{1}{Instituto Nacional de Astrof\'isica, \'Optica y Electr\'onica (INAOE), Apdo.
Postal 51 y 216, 72000 Puebla, Pue. Mexico}
\altaffiltext{2}{Instituto de Astronom\'ia, Universidad Nacional Aut\'onoma de 
M\'exico, Apdo. Postal 70264,04510, M\'exico D. F., Mexico}
\email{gdelgado@astro.unam.mx, mrodri@inaoep.mx}

\begin{abstract}We study the dust present in 56 Galactic planetary nebulae (PNe)
through their iron depletion factors, their C/O abundance ratios (in 51 objects), and the
dust features that appear in their infrared spectra (for 33 objects). Our sample objects
have deep optical spectra of good quality, and most of them also have ultraviolet
observations. We use these observations to derive the iron abundances and the C/O abundance
ratios in a homogeneous way for all the objects. We compile detections of infrared dust
features from the literature and we analyze the available {\it Spitzer}/IRS spectra. Most
of the PNe have C/O ratios below one and show crystalline silicates in their infrared
spectra. The PNe with silicates have C/O~$<1$, with the exception of Cn~1-5. Most of the
PNe with dust features related to C-rich environments (SiC or the 30 $\mu$m feature usually
associated to MgS) have C/O~$\ga0.8$. PAHs are detected over the full range of C/O values,
including 6 objects that also show silicates. Iron abundances are low in all the objects,
implying that more than 90\% of their iron atoms are deposited into dust grains. The range
of iron depletions in the sample covers about two orders of magnitude, and we find that the
highest depletion factors are found in C-rich objects with SiC or the 30 $\mu$m feature in
their infrared spectra, whereas some of the O-rich objects with silicates show the lowest
depletion factors.
\end{abstract}


\section{Introduction}
\label{sec:intro}

Planetary nebulae (PNe) are suitable sites to study the evolution of dust grains 
since their progenitors, asymptotic giant branch (AGB) stars, are among the most efficient
sources of circumstellar dust in the Galaxy \citep{Whittet_03}.
These stars lose large amounts of material and create circumstellar envelopes with both 
low temperatures (below 1500 K in the regions of dust formation) and high densities 
($\sim$ 10$^{13}$ cm$^{-3}$), where dust grains are efficiently formed.

In \citet{DelgadoInglada_09}, we studied iron depletions into dust grains in a sample of 33
Galactic disk PNe. Iron is an important contributor to the mass of dust grains \citep{Sofia_94},
and its gaseous abundance can be used as a proxy for the amount of depletion of 
other elements. We found low iron gaseous abundances in all the objects, suggesting that
more than 90\% of their iron atoms are located in grains. We also found that the range of iron
depletions covers about two orders of magnitude, which may be a consequence of differences in
the formation and evolution of the grains in the different nebulae. We did not find any obvious
correlation between iron abundances and parameters related to the nebular age (such as the
nebular radius or the surface brightness) or to the dominant chemistry in the nebulae (C-rich
or O-rich) that could provide clues on the reasons behind the different depletions.
However, the information available for the studied sample was scarce. 

On the other hand, the dust features that appear in the infrared spectra of PNe provide
information on the composition of their dust grains. In our galaxy, PNe with crystalline
silicates, many of them also with PAHs, are the most numerous \citep{Gutenkunst_08, PC_09,
Stanghellini_12}. Besides, a general correlation has been found between infrared features and
C/O abundance ratios: Magellanic Clouds PNe with silicates have C/O~$<1$, and those with
carbonaceous dust have C/O~$>1$, although for Galactic
PNe the situation is less clear \citep{Casassus_01a,Cohen_05,Stanghellini_07,Waters_98}.

The relative abundances of carbon and oxygen in the envelope of an AGB star depends on the
nucleosynthesis processes occurring in the inner regions, which in turn depends on the mass of
the progenitor star and on its initial metallicity. In our galaxy, we expect stars to be born
with an almost solar C/O ratio (C/O~$\sim0.5$, \citealt{AllendePrieto_02}), that may change
during the evolution of the star depending on its initial mass. Theoretical models \citep[see,
e.g.,][]{Marigo_03, Karakas_09} predict that the less massive progenitors
(M$_{*}$~$\lesssim1.5$--2 M$_\odot$) will  remain O-rich during their whole evolution. Stars
with masses in the range  $\sim$ 2--4 M$_\odot$ may evolve from O-rich to C-rich due to the
third  dredge-up process that enriches the surface of the AGB star with carbon. In the most
massive progenitors (M$_{*}$ $\gtrsim$ 4--5 M$_\odot$), the hot bottom burning process occurs
efficiently and, since it converts carbon into nitrogen via the CN cycle, it can prevent the
formation of a C-rich star.

The value of C/O in a dust-forming stellar envelope defines whether the dust particles will be
C-rich or O-rich. If carbon is more abundant than oxygen, all the oxygen atoms will be trapped
in CO molecules and carbon atoms will be available to form C-rich compounds, such as graphite,
amorphous carbon, or silicon carbide (SiC). If oxygen is more abundant than carbon, all the
carbon atoms will be trapped by CO and the remaining oxygen atoms will form O-rich grains like
silicates and oxides. 

Oxygen rich nebulae could have iron deposited in metallic iron grains, silicates and oxides,
whereas C-rich nebulae are expected to have their iron atoms in the form of metallic grains,
Fe$_3$C, FeSi, FeS, and FeS$_2$ \citep{Whittet_03}. Therefore, iron depletion factors could be
different in C-rich and O-rich PNe. The C-rich or O-rich chemistry may be studied from the C/O
abundance ratios or from the infrared dust features, and we expect both to be related. In
principle, C-rich features such as SiC are expected in PNe with C/O~$>1$
\citep[but see][]{Bond_10} whereas O-rich
features like silicates are expected in PNe with C/O~$<1$. {\it ISO} and {\it Spitzer}
telescopes allow us to detect and identify several dust species in the spectra of PNe and
therefore, to classify PNe as C-rich or O-rich according to the observed features. 

Here, we extend the work performed in \citet{DelgadoInglada_09} by studying a sample of
Galactic PNe that nearly doubles the previous one and also by analyzing possible correlations
between their iron depletions, C/O abundance ratios, and infrared dust features.

The paper is organized as follows. In Section 2 we describe our sample of Galactic PNe, whereas
in Section 3 we present the calculation of electron temperatures and densities for all the
objects. Section 4 contains an extensive discussion of the iron abundance and depletion factor
calculations. Section 5 details the calculation of C/O abundance ratios from collisionally
excited lines (CELs) and recombination lines (RLs), and the comparison between these two
estimates. The compilation of infrared dust features observed in the sample is presented in
Section 6. In Section 7 we discuss our results and study the correlation between iron
abundances, C/O values, and dust features in our sample of PNe. Our conclusions are presented
in Section 8.   

\section{The Sample}
\label{sec:sample}

Our sample consists of 56 Galactic PNe with published spectra of relatively high
spectral resolution and deep enough to detect the lines we need to calculate
physical conditions and iron abundances. The sample includes 28 Galactic PNe studied in
\citet{DelgadoInglada_09}, but we exclude nebulae with electron densities above
25\,000 \cmtres\ since it is difficult to derive reliable physical conditions and chemical
abundances in these objects \citep{DelgadoInglada_09}. 
Besides, we set an upper limit on the $I(\mbox{\ion{He}{2}~}\lambda4686)/I(\mbox{H}\beta)$ 
intensity ratio of 0.6, in order to avoid high excitation PNe for which the iron correction
scheme adopted here may not apply \citep{Rodriguez_05}. 
The final sample of 56 PNe is presented in Table~\ref{tab:1}. 

\begin{deluxetable}{lllll}
\tabletypesize{\scriptsize}
\tablecaption{Physical conditions\label{tab:1}}
\tablewidth{0pt}
\tablehead{
\colhead{Object} & \colhead{\dens\ (\cmtres)}     & \colhead{\temp\nii\ (K)} & \colhead{\temp\oiii\ (K)} & \colhead{Ref.}}
\startdata
\objectname{Cn~1-5}   & $4200^{+1800}_{-1200}$   & $7300\pm200$            & $8800\pm100$  & 1 \\
\objectname{Cn~3-1}    & $7200^{+1800}_{-1200}$   & $7800\pm200$           & $7800\pm400$  & 2 \\	
\objectname{DdDm~1}   & $5300^{+3400}_{-1800}$   & $13100^{+600}_{-800}$   & $12300\pm300$ & 2 \\
\objectname{H~1-41}   & $1600^{+900}_{-500}$     & $9800^{+800}_{-900}$    & $9800\pm200$  & 1 \\
\objectname{H~1-42}   & $8700^{+2600}_{-1700}$   & $11500\pm1200$          & $10200\pm200$ & 1 \\	
\objectname{H~1-50}   & $11200^{+2300}_{-1600}$  & $11000^{+400}_{-500}$   & $11000\pm200$ & 1 \\	
\objectname{Hu~1-1}   & $1800\pm300$             & $11500^{+400}_{-300}$   & $12000\pm300$ & 2 \\	
\objectname{Hu~2-1}   & $9000^{+7500}_{-2900}$   & $12300^{+700}_{-1300}$  & $11000^{+200}_{-300}$ & 2\\
\objectname{IC~418}   & $10300^{+1700}_{-1400}$  & $9400\pm300$            & $8900\pm200$  & 3 \\ 
\objectname{IC~1747}  & $3800^{+800}_{-600}$     & $12400^{+400}_{-500}$   & $10500\pm200$ & 2 \\
\objectname{IC~2165}  & $4500^{+600}_{-500}$     & $13100\pm500$           & $14600\pm400$ & 4 \\
\objectname{IC~3568}  & $2100^{+700}_{-500}$    & $18800^{+1200}_{-2900}$  & $11500\pm300$ & 5 \\	
\objectname{IC~4191}  & $14800^{+2400}_{-2300}$ & $11300\pm500$            & $10300\pm200$ & 6 \\
\objectname{IC~4406}  & $2400^{+400}_{-300}$    & $10200\pm300$            & $10000\pm200$ & 6 \\
\objectname{IC~4593}  & $1700^{+700}_{-500}$    & $9800^{+900}_{-1000}$    & $8500^{+200}_{-300}$ & 7 \\
\objectname{IC~4699}  & $2500^{+1700}_{-900}$   & $19500^{+500}_{-4600}$   & $11800\pm300$ & 1 \\
\objectname{IC~4846}  & $10100^{+4000}_{-2400}$ & $12100^{+2000}_{-2100}$  & $10600\pm400$ & 8 \\  
\objectname{IC~5217}  & $5100^{+1300}_{-1000}$  & $13700^{+3400}_{-2800}$  & $10800\pm400$ & 9 \\ 
\objectname{JnEr~1}   & $400^{+600}_{-300}$     & $10300^{+900}_{-800}$    & $12700^{+2000}_{-1000}$ & 7 \\
\objectname{M~1-20}   & $12100^{+4900}_{-3200}$ & $11000^{+600}_{-700}$    & $9900\pm200$ & 1 \\	
\objectname{M~1-42}   & $1400^{+300}_{-200}$    & $9100\pm200$             & $10000\pm200$ & 10 \\
\objectname{M~1-73}   & $4300^{+1600}_{-900}$   & $8800^{+200}_{-300}$     & $7300\pm100$ & 2 \\
\objectname{M~2-4}    & $8700^{+3100}_{-2200}$  & $10000^{+400}_{-500}$    & $8600\pm100$ & 1\\	
\objectname{M~2-6}    & $10300^{+6200}_{-3500}$ & $10600^{+600}_{-800}$    & $10100\pm200$ & 1\\	
\objectname{M~2-27}   & $13600^{+3500}_{-2300}$ & $8800^{+300}_{-400}$     & $8300\pm100$ & 1 \\	
\objectname{M~2-31}   & $7700^{+2300}_{-1500}$  & $11400^{+400}_{-500}$    & $9900\pm200$ & 1 \\	
\objectname{M~2-33}   & $3100^{+3600}_{-1600}$  & $9200^{+1000}_{-1300}$   & $8000\pm100$ & 1 \\	
\objectname{M~2-36}   & $4700^{+700}_{-600}$   & $9200^{+400}_{-100}$      & $8400\pm100$ & 10 \\
\objectname{M~2-42}  & $4800^{+2400}_{-1600}$  & $10200^{+400}_{-500}$     & $8500\pm100$ & 1 \\	
\objectname{M~3-7}  & $4700^{+2400}_{-1500}$  & $8600\pm300$               & $7700\pm100$ & 1 \\	
\objectname{M~3-29}   & $800^{+400}_{-200}$     & $9000^{+600}_{-700}$     & $9200\pm200$ & 1 \\	
\objectname{M~3-32}   & $2900^{+800}_{-600}$    & $17400\pm2500$           & $8900\pm200$ & 1 \\	
\objectname{MyCn~18}  & $9400^{+1700}_{-1300}$ & $9800\pm300$              & $7400\pm100$  & 6 \\
\objectname{NGC~40}   & $1300\pm200$           & $8600\pm200$              & $10600\pm200$ & 5 \\
\objectname{NGC~2392} & $2200^{+1900}_{-1000}$ & $12700^{+1600}_{-1300}$   & $14600^{+900}_{-1000}$ & 7\\
\objectname{NGC~3132} & $700\pm200$           & $9700^{+200}_{-300}$       & $9600\pm200$  & 6 \\
\objectname{NGC~3242} & $2400\pm400$           & $12400\pm400$             & $11900\pm300$ & 6 \\ 	
\objectname{NGC~3587} & $2400^{+1900}_{-1300}$ & $10900\pm1000$            & $11600\pm500$ & 7 \\
\objectname{NGC 3918} & $7100^{+1100}_{-900}$ & $10800^{+300}_{-400}$      & $12700\pm300$ & 6\\
\objectname{NGC~5882} & $5100^{+700}_{-500}$   & $10600\pm300$             & $9400\pm200$  & 6 \\ 
\objectname{NGC~6153} & $3900^{+500}_{-400}$   & $10500\pm300$             & $9200\pm200$  & 11 \\
\objectname{NGC~6210} & $5800^{+4500}_{-2100}$ & $11000^{+400}_{-300}$     & $9600\pm200$  & 5 \\
\objectname{NGC~6439} & $6000^{+800}_{-700}$   & $9700\pm300$              & $10400\pm200$ & 1 \\
\objectname{NGC~6543} & $5800^{+2000}_{-1400}$ & $10000^{+500}_{-600}$     & $7900\pm200$  & 12 \\  
\objectname{NGC~6565} & $1600\pm300$           & $10600\pm300$             & $10400\pm200$ & 1 \\
\objectname{NGC~6572} & $17900^{+4300}_{-2700}$& $12000^{+600}_{-700}$     & $10400\pm200$ & 5 \\  
\objectname{NGC~6620} & $2900\pm400$           & $9000\pm200$              & $9600\pm200$ & 1  \\
\objectname{NGC~6720} & $700\pm200$            & $10600\pm300$             & $10700^{+300}_{-200}$  & 5 \\   
\objectname{NGC 6741} & $6700\pm800$           & $10800^{+400}_{-300}$     & $12600\pm300$ & 5 \\
\objectname{NGC~6803} & $9300^{+1200}_{-900}$  & $10700\pm300$             & $9700\pm200$  & 2 \\ 
\objectname{NGC 6818} & $2300\pm300$           & $11400\pm400$             & $13400^{+400}_{-300}$ & 6 \\
\objectname{NGC~6826} & $2100\pm300$           & $10600\pm500$             & $9400\pm200$  & 5 \\  
\objectname{NGC~6884} & $8900^{+1300}_{-1000}$ & $11600\pm400$             & $11100\pm300$ & 5 \\
\objectname{NGC~7026} & $7500^{+1100}_{-800}$  & $9700\pm300$              & $9300\pm200$ & 2 \\
\objectname{NGC 7662} & $3000\pm400$           & $13000^{+500}_{-400}$     & $13400\pm400$  & 5 \\
\objectname{Vy~2-1}   & $5400^{+2600}_{-1800}$ & $9300^{+100}_{-700}$      & $7900\pm100$  & 1 
\enddata
\tablerefs{{\it Line intensities from:} (1) \citet{Wang_07}, (2) \citet{Wesson_05}, (3) \citet{Sharpee_03}, 
(4) \citet{Hyung_94}, (5) \citet{Liu_04b}, (6) \citet{Tsamis_03}, (7) \citet{DelgadoInglada_09}, 
(8) \citet{Hyung_01b}, (9) \citet{Hyung_01a}, (10) \citet{Liu_01}, (11) \citet{Liu_00}, 
(12) \citet{Wesson_04}.}
\end{deluxetable}

\section{Physical conditions}
\label{sec:phys_cond}

We derive electron densities and temperatures, \dens\ and \temp, 
using the routine {\it temden} from the package {\it nebular} 
\citep{Shaw_95} in {\sc IRAF}\footnote{{\sc IRAF} is distributed by the 
National Optical Astronomy Observatories, which are operated by the Association 
of Universities for Research in Astronomy, Inc., under cooperative agreement 
with the National Science Foundation.}.
For each object, we calculate two temperatures from the observed intensity 
ratios \nii\ $\lambda5755/(\lambda6548+\lambda6584)$ and 
\oiii\ $\lambda4363/(\lambda4959+\lambda5007)$ to characterize the low- and 
high-ionization regions, respectively. We also derive an average density from 
the available diagnostic ratios among the following: 
\oii\ $\lambda3726/\lambda3729$, \sii\ $\lambda6716/\lambda6731$, 
\cliii\ $\lambda5518/\lambda5538$, and \ariv\ $\lambda4711/\lambda4740$. 
We use the default atomic data in {\sc IRAF} except for Cl$^{++}$ and
O$^+$: for these ions we find densities in better 
agreement with those obtained from the other diagnostic ratios using
the collision strengths of \citet{Krueger_70} and \citet{Tayal_07} and
the transition probabilities of \citet{Mendoza_82} and \citet{Zeippen_82}. 

\citet{Rubin_86} found that a significant fraction of the intensity of the
\nii\ $\lambda$5755 line can arise from recombinations of N$^{++}$.
If this effect is not taken into account, the value of
\temp\nii\ can be overestimated in those objects where N$^{++}$ is an 
important contributor to the total abundance of nitrogen. The contribution of
recombination to \nii\ $\lambda$5755 can be estimated using the 
expression derived by \citet{Liu_00}, which depends on the N$^{++}$ abundance.
This ionic abundance can be obtained either from infrared/ultraviolet
collisionally excited lines or from optical recombination lines, leading
in each case to different results. Thus, the correction is somewhat uncertain
and we do not apply it here. However, we estimate an upper limit of the effect
it could have on our results by using the highest possible values for the N$^{++}$
abundances, which are those implied by recombination lines. We find that in many
of our objects \temp\nii\ would change by less than 10\%. The objects with the
largest differences in \temp\nii\ are DdDm~1, IC~3568, IC~4699, M~3-32, and
NGC~3242, where they reach 30--50\%. For these objects, this implies
values of Fe/O up to 0.4 dex lower or up to 0.2 dex higher, depending on
which ionization correction approach we use (see \S~\ref{sec:tot_ab}). Hence,
the iron abundances derived for these 5 PNe have this additional amount of uncertainty. 

We calculate the uncertainties of the physical conditions and the ionic and total
abundances of O and Fe using Montecarlo simulations, which assume that the errors
in the line intensities follow Gaussian distributions. The line intensity errors
are those given in the references listed in Table~\ref{tab:1}. Each Montecarlo run
samples the Gaussian distributions of all the lines in order to provide new values for
\temp, \dens, and all the ionic and total abundances described in \S~\ref{sec:ionic_ab}
and \S~\ref{sec:tot_ab} below.
The errors listed for the calculated quantities define a confidence interval of 68$\%$ 
(equivalent to one standard deviation in a Gaussian distribution). 

Table~\ref{tab:1} presents the physical conditions and their associated
uncertainties for our sample PNe. The values are similar to the ones obtained in 
\citet{DelgadoInglada_09} for the objects in common. The highest differences are
for the values of \dens, and there are two reasons for that. First, we use here one more 
diagnostic ratio for the density, [\ion{O}{2}] $\lambda3726/\lambda3729$. Second,
whereas \citet{DelgadoInglada_09} derive the final \dens\ as the weighted mean
of the values implied by the available diagnostic ratios, here the final \dens\ 
is the median value of the density distribution obtained from averaging the \dens\ 
values calculated in each run of the Montecarlo simulation. One of the objects with
significant differences is NGC~3587, where \citet{DelgadoInglada_09} used the
[\ion{S}{2}] density diagnostic and found \dens\ $<100$ \cmtres, whereas here we
also use the [\ion{O}{2}] diagnostic and find \dens $= 2400$ \cmtres. This change in
\dens\ implies a change in \temp\nii\ from 9800 K to 10900 K. In general, \temp\
differences are smaller, around 100 K in most of the objects.
The new values derived for the physical conditions imply oxygen abundances that are
within $\pm$0.15 dex of the old values. These differences illustrate the abundance
uncertainties introduced by the approach chosen to perform the calculations.

\section{Iron depletion factors}
\label{sec:depletion}

\subsection{Ionic abundances}
\label{sec:ionic_ab}

The total iron abundances of the sample PNe are calculated from the Fe$^{++}$ abundance
(and Fe$^{+}$ in some cases) along with ionization correction factors (ICFs) based on
the ionic and total oxygen abundances. On the other hand, the ICF we use for oxygen is
based on the He$^{+}$ and He$^{++}$ abundances. Hence, we derive all these ionic
abundances using the densities and temperatures from Table~\ref{tab:1}, in particular
\temp\nii\ for O$^{+}$, Fe$^{+}$, and Fe$^{++}$, and 
\temp\oiii\ for O$^{++}$, He$^{+}$, and He$^{++}$. 

The values of O$^{+}$/H$^{+}$ and O$^{++}$/H$^{+}$ are calculated using the intensities of
the [\ion{O}{2}] $\lambda\lambda3726+29$ and [\ion{O}{3}] $\lambda\lambda4959,5007$ lines
with respect to H$\beta$ and the routine {\it ionic} of the package {\it nebular} in
{\sc IRAF} with the changes in atomic data mentioned in \S~\ref{sec:phys_cond}. 

The values of He$^{+}$/H$^{+}$ are based on \ion{He}{1} $\lambda$6678 and the calculations 
of \citet{Benjamin_99}. In DdDm~1 and IC~2165 we replace our computed He$^{+}$ 
abundances (He$^{+}$/H$^{+}=0.041$ and 0.042) with the average of the He$^{+}$ abundances
derived from the \ion{He}{1} $\lambda$$\lambda$4471, 5876 lines by \citet{Hyung_94} and
\citet{Wesson_05}, respectively. These authors suggest that the low values implied by the
$\lambda$6678 line could be due to underlying absorption.

The values of He$^{++}$/H$^{+}$ are derived using \ion{He}{2} $\lambda$4686 and the 
emissivities of \citet{Storey_95}. The \ion{H}{1} emissivities are also from \citet{Storey_95}. 

We estimate the Fe$^{+}$ abundances for eleven PNe using the emissivities of
\citet{Bautista_96} and [\ion{Fe}{2}] $\lambda8616$, one of the few [\ion{Fe}{2}] lines
which are not affected by fluorescence \citep{Verner_00}. In some cases where this line
is not available we use instead [\ion{Fe}{2}] $\lambda7155$ and
$I(\lambda8616)/I(\lambda7155) \sim 1$,
as found by \citet{Rodriguez_96} in several \ion{H}{2} regions. 
The values of Fe$^{+}$/Fe$^{++}$ for these eleven PNe go from 0.06 (M~3-7) to 0.80 
(NGC~6565 and NGC~6741). Only five of the sample PNe would require a value for the Fe$^{+}$
abundance in the ionization correction scheme described below, in \S~\ref{sec:tot_ab}.
The two of them where it is available show that its contribution is not
likely to be critical, since it only increases the value of Fe/H by 0.08 dex (in NGC~40)
and 0.17 dex (in IC~418). Some of the calculated values of Fe$^{+}$/H$^{+}$ are also used
in the calculations described in \S~\ref{withoutICF}.

The Fe$^{++}$ ionic abundances are derived by solving the equations of statistical 
equilibrium \citep{Osterbrock_06} for a 34-level model atom, using the collision
strengths of \citet{Zhang_96} and the transition probabilities of \citet{Quinet_96}. 
We used PyNeb \citep{Luridiana_12} to compare the results obtained with these atomic 
data with those implied by the recent calculations of \citet{Bautista_10}. In the three
nebulae with the largest numbers of [\ion{Fe}{3}] lines (Cn~1-5, M~2-4, and M~2-6), we find
differences lower than 5\% in the final Fe$^{++}$ abundances, but the older atomic
data lead to a better agreement in the Fe$^{++}$ abundances derived from all the 
available [\ion{Fe}{3}] lines, with standard deviations of 32\%/39\% for the old/new atomic
data (Cn~1-5), 12\%/25\%(M~2-4), and 23\%/25\% (M~2-6). We performed a similar comparison
using 14 out of the 15 [\ion{Fe}{3}] lines measured by \citet{Esteban_04} in the Orion
Nebula (we excluded the line at 4926 \AA\ because it leads to much higher abundances with
both sets of atomic data). We find that the new atomic data lead to a Fe$^{++}$ abundance
that is 15\% lower, whereas the standard deviation changes from 15\% to 30\%. Hence, we
perform our calculations using the older set of atomic data.

We use up to 9 [\ion{Fe}{3}] lines for each object, excluding those suspected to be
contaminated with nearby recombination lines.  
For the eight PNe in the sample with no [\ion{Fe}{3}] lines in their spectra, we estimate
upper limits to their Fe$^{++}$ abundances using the intensities reported for 
\ion{C}{4} $\lambda$4658 or \ion{O}{2} $\lambda$4661. 
These lines are close in wavelength to [\ion{Fe}{3}] $\lambda$4658, the brightest
[\ion{Fe}{3}] line for the physical conditions of our PNe. We use \ion{C}{4}
$\lambda$4658 for IC~4406, IC~1747, JnEr~1, M~1-42, 
M~3-29, and NGC~7026, and \ion{O}{2} $\lambda$4661 for H~1-41 and M~3-32. 
Some PNe show lines from higher iron ionization states; this will be discussed in 
\S~\ref{withoutICF}.

Table~\ref{tab:2} presents all the computed ionic abundances and their uncertainties. The
last column shows the number of [\ion{Fe}{3}] lines used to derive the Fe$^{++}$ abundance
in each object. The PNe where only upper limits are available are distinguished  either
with ``1?'', when the \ion{C}{4} $\lambda$4658 line was used (note that this line could be
a misidentification of [\ion{Fe}{3}] $\lambda$4658), or with ``--'' when \ion{O}{2}
$\lambda$4661 was used. 

\begin{deluxetable*}{lllllllc}
\tabletypesize{\small}
\tablecaption{Ionic Abundances: \{X$^{+i}$\} = 12 + $\log$ (X$^{+i}$/H$^{+}$)\label{tab:2}}
\tablewidth{0pt}
\tablehead{
\colhead{PN} & \colhead{\{He$^{+}$\}} & \colhead{\{He$^{++}$\}} & \colhead{\{O$^{+}$\}} & \colhead{\{O$^{++}$\}} & 
\colhead{\{Fe$^{+}$\}} & \colhead{\{Fe$^{++}$\}} & \colhead{Number of} \\
\colhead{} & \colhead{$\lambda$6678} & \colhead{$\lambda$4686} & \colhead{$\lambda\lambda$3726,3729} & \colhead{$\lambda\lambda$4959.5007} 
& \colhead{$\lambda$8616} & \colhead{average} & \colhead{[\ion{Fe}{3}] lines\tablenotemark{a}}}
\startdata
Cn~1-5   & $11.08\pm0.02$ & --                     & $8.36\pm0.11$              & $8.68\pm0.04$          & 5.23 & $6.29^{+0.05}_{-0.07}$ & 8 \\
Cn~3-1	 & $10.67\pm0.02$ & $7.52^{+0.11}_{-0.15}$ & $8.68^{+0.12}_{-0.09}$       & $7.26^{+0.11}_{-0.08}$   &   --   & $5.57^{+0.06}_{-0.05}$ & 3 \\	
DdDm~1\tablenotemark{b}   & $10.95\pm0.02$ & --                     &  $7.38^{+0.31}_{-0.14}$    & $7.91^{+0.05}_{-0.04}$ & --   & $5.76^{+0.08}_{-0.04}$  & 4\\
H~1-41   & $10.96\pm0.02$ & $10.35\pm0.02$         & $7.27^{+0.19}_{-0.14}$     & $8.50\pm0.04$          & --   & $<4.85$ & --\\	
H~1-42   & $11.05\pm0.02$ & $8.83\pm0.02$          & $7.10^{+0.23}_{-0.18}$     & $8.54\pm0.04$          & 4.50 & $4.67^{+0.17}_{-0.14}$ & 3 \\	
H~1-50   & $10.99\pm0.02$ & $10.04\pm0.02$         & $7.50^{+0.12}_{-0.10}$     & $8.62\pm0.04$          & --   & $4.54^{+0.10}_{-0.11}$ & 1 \\	
Hu~1-1	 & $10.93\pm0.02$ & $10.18\pm0.02$         & $8.00^{+0.06}_{-0.05}$       & $8.40\pm0.04$          &   --   & $4.25^{+0.16}_{-0.22}$ & 1 \\	
Hu 2-1   & $10.83\pm0.02$ & $8.39^{+0.03}_{-0.01}$ & $7.44^{+0.34}_{-0.07}$     & $8.19^{+0.05}_{-0.04}$ & --   & $4.88^{+0.05}_{-0.08}$  & 4 \\
IC~418	 & $10.97\pm0.02$ & --                     & $8.34^{+0.07}_{-0.12}$       & $8.06\pm0.04$          &   3.89 & $4.20^{+0.06}_{-0.05}$ & 5 \\ 
IC~1747	 & $11.01\pm0.02$ & $10.04\pm0.02$         & $7.07^{+0.09}_{-0.06}$       & $8.56\pm0.04$          & --   & $<4.34$              & 1? \\
IC~2165\tablenotemark{b}  & $10.74\pm0.02$ & $10.73\pm0.02$         &  $6.80\pm0.06$             & $8.10\pm0.04$          & 3.64 & $4.58^{+0.06}_{-0.07}$& 3\\
IC~3568	 & $10.96\pm0.02$ & $9.02\pm0.02$          & $5.69^{+0.24}_{-0.06}$       & $8.36\pm0.05$          & --   & $3.90^{+0.17}_{-0.10}$ & 2 \\	
IC~4191  & $11.04\pm0.02$ & $10.08\pm0.02$         & $7.51^{+0.12}_{-0.11}$       & $8.64\pm0.04$          & --   & $4.38^{+0.10}_{-0.11}$ & 1 \\
IC~4406  & $10.97\pm0.02$ & $10.08\pm0.02$         & $8.28\pm0.06$              & $8.58\pm0.04$          & --   & $<4.58$              & 1? \\
IC~4593  & $11.00\pm0.02$ & $8.53^{+0.05}_{-0.06}$ & $7.39^{+0.24}_{-0.16}$       & $8.54\pm0.06$          & --   & $5.39^{+0.16}_{-0.13}$ & 4 \\
IC~4699  & $10.92\pm0.02$ & $10.26\pm0.02$         & $6.14^{+0.36}_{-0.03}$     & $8.40\pm0.04$          & --   & $3.90^{+0.27}_{-0.13}$ & 1 \\
IC~4846  & $10.96\pm0.02$ & $8.68^{+0.11}_{-0.15}$ & $7.03^{+0.40}_{-0.26}$       & $8.48\pm0.06$          & --   & $4.54^{+0.26}_{-0.19}$ & 2 \\  
IC~5217  & $10.84^{+0.08}_{-0.10}$ & $9.95\pm0.04$ & $6.59^{+0.41}_{-0.28}$       & $8.63\pm0.07$          & --   & $4.61^{+0.27}_{-0.30}$ & 1 \\ 
JnEr~1   & $11.29^{+0.08}_{-0.09}$ & $10.25\pm0.04$& $8.40^{+0.19}_{-0.18}$       & $7.83^{+0.16}_{-0.13}$   & --   & $<5.49$              & 1?\\
M~1-20   & $10.99\pm0.02$ & $7.61^{+0.12}_{-0.16}$  & $7.46^{+0.21}_{-0.17}$     & $8.53\pm0.04$          & --   & $4.39^{+0.16}_{-0.17}$ & 1 \\	
M~1-42   & $11.22\pm0.02$ & $10.04\pm0.02$         & $7.61^{+0.16}_{-0.05}$     & $8.36^{+0.05}_{-0.04}$ & --   & $<5.53$  & 1?\\
M~1-73   & $11.02\pm0.02$ & $8.99\pm0.02$          & $8.13^{+0.13}_{-0.09}$     & $8.52\pm0.05$          & --   & $5.42^{+0.07}_{-0.06}$ & 2 \\
M~2-4    & $11.08\pm0.02$ & --                     & $7.82^{+0.18}_{-0.13}$     & $8.66\pm0.04$          & --   & $5.49\pm0.07$          & 9 \\	
M~2-6    & $11.05\pm0.02$ & $8.93\pm0.02$          & $7.70^{+0.26}_{-0.18}$     & $8.36\pm0.04$          & --   & $5.34^{+0.11}_{-0.08}$ & 8 \\	
M~2-27   & $11.13\pm0.02$ & $8.84\pm0.02$          & $7.83^{+0.16}_{-0.12}$     & $8.82\pm0.04$          & --   & $5.51^{+0.09}_{-0.06}$ & 2\\	
M~2-31   & $11.10\pm0.02$ & --                     & $7.30^{+0.13}_{-0.09}$     & $8.62\pm0.04$          & --   & $5.55^{+0.14}_{-0.15}$ & 1\\	
M~2-33   & $11.01\pm0.02$ & $8.92\pm0.02$          & $7.34^{+0.41}_{-0.19}$     & $8.67\pm0.04$          & --   & $5.30^{+0.24}_{-0.16}$ & 5\\	
M~2-36   & $11.01\pm0.02$ & $9.01\pm0.02$          &  $7.64^{+0.10}_{-0.07}$    & $8.71\pm0.04$          & --   & $4.37^{+0.13}_{-0.16}$  & 1\\
M~2-42   & $11.05\pm0.02$ & $8.45^{+0.08}_{-0.10}$  & $7.37^{+0.15}_{-0.10}$     & $8.70\pm0.04$          & --   & $5.16^{+0.10}_{-0.10}$ & 2\\	
M~3-7 	 & $11.05\pm0.02$ & $9.20\pm0.02$          & $7.83^{+0.16}_{-0.11}$     & $8.62\pm0.04$          &   4.56 & $5.75^{+0.09}_{-0.08}$ & 2 \\	
M~3-29   & $10.98\pm0.02$ & --                     & $7.68^{+0.17}_{-0.14}$     & $8.40\pm0.04$          & --   & $<5.31$ & 1?\\	
M~3-32   & $11.09\pm0.02$ & $10.02\pm0.02$         & $6.18^{+0.20}_{-0.14}$     & $8.56\pm0.04$          & --   & $<4.81$ & -- \\	
MyCn~18  & $10.94\pm0.02$ & $8.66\pm0.02$          & $7.78^{+0.11}_{-0.08}$     & $8.50\pm0.04$          & --   & $5.47\pm0.05$          & 6 \\
NGC~40   & $10.80\pm0.02$ & $7.56^{+0.11}_{-0.14}$ & $8.61^{+0.06}_{-0.05}$     & $7.06\pm0.05$          & 4.86 & $5.55^{+0.05}_{-0.04}$ & 5 \\
NGC~2392 & $10.89^{+0.08}_{-0.10}$ & $10.46\pm0.04$& $7.40^{+0.23}_{-0.24}$     & $8.06^{+0.09}_{-0.07}$ & --   & $5.63^{+0.13}_{-0.12}$ & 5 \\	
NGC~3132 & $11.04\pm0.02$ & $9.51\pm0.02$          & $8.39\pm0.05$              & $8.51\pm0.04$          & --   & $5.19^{+0.07}_{-0.09}$ & 6  \\
NGC~3242 & $10.90\pm0.02$ & $10.33\pm0.02$         & $6.48^{+0.06}_{-0.05}$     & $8.41^{+0.05}_{-0.04}$ & --   & $4.03^{+0.07}_{-0.08}$ & 2 \\ 	
NGC~3587 & $10.91^{+0.07}_{-0.09}$&$10.15^{+0.02}_{-0.03}$&$8.01^{+0.21}_{-0.18}$& $8.26^{+0.08}_{-0.07}$& --   & $6.09^{+0.22}_{-0.37}$ & 1 \\
NGC 3918 & $10.84\pm0.02$ & $10.55\pm0.02$         &  $7.71^{+0.10}_{-0.08}$    & $8.43\pm0.04$          & --   & $4.28^{+0.09}_{-0.10}$  & 3\\
NGC~5882 & $11.02\pm0.02$ & $9.35\pm0.02$          & $6.91\pm0.06$              & $8.65\pm0.04$          & --   & $4.74^{+0.07}_{-0.06}$ & 6 \\ 
NGC~6153 & $11.05\pm0.02$ & $10.05\pm0.02$         & $7.16\pm0.06$              & $8.61\pm0.04$          & --   & $4.56^{+0.08}_{-0.09}$ & 2 \\
NGC~6210 & $11.02\pm0.02$ & $9.28\pm0.02$          & $7.26^{+0.10}_{-0.09}$     & $8.53^{+0.06}_{-0.05}$ & --   & $4.65^{+0.08}_{-0.07}$ & 1 \\
NGC~6439 & $11.09\pm0.02$ & $10.31\pm0.02$         & $7.73\pm0.07$              & $8.58\pm0.04$          & --   & $4.96^{+0.09}_{-0.10}$ & 2\\
NGC~6543 & $11.05\pm0.02$ & --                     & $7.25^{+0.15}_{-0.13}$     & $8.74\pm0.06$          & --   & $4.93\pm0.08$        & 6  \\  
NGC~6565 & $11.00\pm0.02$ & $10.19\pm0.02$         & $8.06\pm0.05$              & $8.56\pm0.04$          & 4.90 & $5.40\pm0.06$          & 7\\
NGC~6572 & $11.01\pm0.02$ & $8.53\pm0.02$          & $7.41^{+0.20}_{-0.10}$     & $8.56^{+0.06}_{-0.04}$ & --   & $4.53\pm0.08$        & 7 \\  
NGC~6620 & $11.07\pm0.02$ & $10.33\pm0.02$         & $8.20\pm0.06$              & $8.69\pm0.04$          & 5.31 & $5.05^{+0.12}_{-0.16}$ & 1\\
NGC~6720 & $10.96\pm0.02$ & $10.25\pm0.02$         & $8.24^{+0.06}_{-0.05}$     & $8.46\pm0.05$          & 4.26 & $4.77\pm0.06$        & 4 \\   
NGC 6741 & $10.89\pm0.02$ & $10.49\pm0.02$         &  $8.10^{+0.19}_{-0.11}$    & $8.39\pm0.05$          & 5.58 & $5.68^{+0.07}_{-0.05}$& 7\\
NGC~6803 & $11.04\pm0.02$ & $9.56\pm0.02$          & $7.38\pm0.08$              & $8.63\pm0.05$          & --   & $4.87^{+0.06}_{-0.05}$ & 3 \\ 
NGC 6818 & $10.73\pm0.02$ & $10.72\pm0.02$         &  $7.36^{+0.07}_{-0.06}$    & $8.35\pm0.04$          & --   & $4.68\pm0.06$ & 1\\
NGC~6826 & $11.00\pm0.02$ & $7.34^{+0.11}_{-0.15}$ &$6.99\pm0.09$               & $8.50\pm0.05$          & --   & $4.69\pm0.08$        & 3 \\  
NGC~6884 & $10.87\pm0.02$ & $10.19\pm0.02$         & $7.16^{+0.09}_{-0.07}$     & $8.53\pm0.05$          & 4.04 & $4.72\pm0.06$        & 6 \\
NGC~7026 & $11.04\pm0.02$ & $10.11\pm0.02$         & $7.76^{+0.06}_{-0.05}$     & $8.60\pm0.05$          & --   & $<4.67$ & 1? \\
NGC 7662 & $10.82\pm0.02$ & $10.58\pm0.02$         &  $6.28\pm0.06$             & $8.23^{+0.06}_{-0.05}$ & --   & $4.40\pm0.06$ & 4 \\
Vy~2-1   & $11.11\pm0.02$ & $8.71^{+0.08}_{-0.10}$  & $7.90^{+0.23}_{-0.06}$     & $8.72\pm0.04$          & --   & $4.86^{+0.16}_{-0.07}$ & 2
\enddata
\tablenotetext{a}{``1?'': upper limit for Fe$^{++}$/H$^{+}$ from \ion{C}{4} $\lambda$4658, 
``--'': upper limit for Fe$^{++}$/H$^{+}$ from \ion{O}{2} $\lambda$4661.}
\tablenotetext{b}{He$^+$/H$^{+}$ derived using \ion{He}{1} $\lambda\lambda$4471, 5876
(see the text for details).}
\end{deluxetable*}

\subsection{Total abundances with an ICF}
\label{sec:tot_ab}

The absence of \ion{He}{2} lines in seven PNe of the sample indicates that their He$^{++}$ 
abundances are negligible, and since the ionization potentials of O$^{++}$ and He${^+}$ are 
very similar (54.93 eV and 54.42 eV, respectively), we do not expect an important amount of 
O$^{3+}$ in these nebulae. 
Therefore, in these low-ionization PNe, we simply add the ionic abundances of O$^{+}$ and
O$^{++}$ to calculate the oxygen abundance. For the other PNe we use the expression

\begin{equation}\label{eq:ICF_O}
\frac{\mbox{O}}{\mbox{H}} = \left( \frac{\mbox{O}^{+} + \mbox{O}^{++}}{\mbox{H}^{+}}\right)\left(\frac{\mbox{He}^{+} + \mbox{He}^{++}}{\mbox{He}^{+}}\right) 
\end{equation}

\noindent \citep{Peimbert_71}. 
The ICFs of oxygen in most of our sample PNe have values below 1.2. The exceptions are the
PNe with the highest $I(\mbox{\ion{He}{2}~}\lambda4686)/I(\mbox{H}\beta)$ ratios: IC~2165
(ICF = $1.98$), NGC~3242 (1.28), NGC~3918 (1.51), NGC~6741 (1.40), NGC~6818 (1.97), and
NGC~7662 (1.57). The differences between the values of O/H derived from
equation~(\ref{eq:ICF_O}) and those derived with the ICF suggested in
\citet{DelgadoInglada_14} are lower than 0.02 dex for most of the nebulae, and reach a
maximum value of 0.13 dex for IC~2165 and NGC~6818. The differences are small because both
ICFs lead to similar corrections for low ionization PNe, with
$\mbox{He}^{++}/(\mbox{He}^{+}+\mbox{He}^{++})<0.5$.

In general, our values of O/H agree within 0.1 dex with those derived in the papers listed 
in Table~\ref{tab:1}. Ten PNe show differences of up to 0.4 dex, and are those with the
highest differences in the adopted physical conditions and in the correction applied for
the unobserved ions. Again, this illustrates the uncertainties introduced by the
choices one has to make in order to calculate chemical abundances.

The [\ion{Fe}{3}] lines are usually the brightest iron lines in photoionized nebulae.
Therefore, total iron abundances are generally calculated using $\mbox{Fe}^{++}$ abundances 
and ICFs derived from photoionization models. However the iron abundances computed in 
this way differ from the ones obtained by adding up the ionic abundances of Fe$^{+}$,
Fe$^{++}$, and Fe$^{+3}$ in some \ion{H}{2} regions and low-excitation PNe. 
\citet{Rodriguez_05} studied the uncertainties involved in the calculations and proposed
that the iron abundances can be constrained using two ICFs, one derived 
from photoionization models: 
\begin{equation}\label{eq:1}
\frac{\mbox{Fe}}{\mbox{O}}=0.9\left(\frac{\mbox{O}^{+}}{\mbox{O}^{++}}\right)^{0.08}\frac{\mbox{Fe}^{++}}{\mbox{O}^{+}},
\end{equation}
\noindent and the other computed from observational data:
\begin{equation}\label{eq:2}
\frac{\mbox{Fe}}{\mbox{O}}=1.1\left(\frac{\mbox{O}^{+}}{\mbox{O}^{++}}\right)^{0.58}\frac{\mbox{Fe}^{++}}{\mbox{O}^{+}}.
\end{equation}

For low ionization objects, with $\log(\mbox{O}^+/\mbox{O}^{++})\geq-0.1$, 
equation~(\ref{eq:2}) should be replaced by
\begin{equation}\label{eq:3}
\frac{\rm {Fe}}{\rm {O}}=\frac{\mbox{Fe}^+ + \mbox{Fe}^{++}}{\mbox{O}^+},
\end{equation}
since $\mbox{Fe}^{++}$ and $\mbox{O}^+$ will dominate the total abundances of Fe and O.
This scheme provides three values for the iron abundance, each of them corresponding
to one of the three possible explanations suggested by \citet{Rodriguez_05} for the iron
abundance discrepancy. The Fe/O values obtained from equation~(2) should be used if the
collision strengths for Fe$^{+3}$ are wrong. If the collision strengths for Fe$^{++}$ are
incorrect, the best values are those obtained from equation~(2) lowered by 0.3 dex. And
third, if the recombination coefficient or the rate of charge-exchange reaction for
Fe$^{+3}$ are wrong, the ICF from equation~(3) should be used. If there is a combination of
causes for the discrepancy, the real iron abundance for each object will be intermediate
between the two extremes given by the three values. Since equations~(2) and (3) lead to
different abundance distributions, and since we want to explore possible correlations with
other parameters, we keep the individual abundances obtained from each ionization
correction scheme, instead of just providing the average value.

Table~\ref{tab:3} shows the final abundances of oxygen and iron. The results shown for
iron are the values obtained from equations~(\ref{eq:1}) and (\ref{eq:2}). The third set
of values, needed to constrain the iron abundances in the manner described above, can be
obtained by subtracting $\sim 0.3$~dex to the values in column~(3). Taking into account
these three values, we constrain the iron abundances to the abundance range shown in
column~(5).

\begin{deluxetable}{lllll}
\tabletypesize{\footnotesize}
\tablecaption{Total Abundances:  \{X\} = 12 + $\log$ (X/H)\label{tab:3}}
\tablewidth{0pt}
\tablehead{
\colhead{PN} & \colhead{\{O\}} & \colhead{\{Fe\}\tablenotemark{a}} &
\colhead{\{Fe\}\tablenotemark{b}} & \colhead{$\Delta$\{Fe\}}}
\startdata
Cn~1-5   & $8.85^{+0.06}_{-0.05}$ &  $6.71^{+0.07}_{-0.09}$  & $6.63^{+0.05}_{-0.07}$ & 6.4--6.7 \\
Cn~3-1	 & $8.69^{+0.11}_{-0.08}$ & $5.66^{+0.06}_{-0.05}$   & $5.59^{+0.06}_{-0.05}$ & 5.4--5.7 \\
DdDm 1   & $8.02^{+0.11}_{-0.05}$ &  $6.31^{+0.12}_{-0.16}$  & $6.13^{+0.08}_{-0.06}$ & 6.0--6.3 \\
H~1-41   & $8.62^{+0.05}_{-0.03}$ &  $<6.05$                 & $<5.52$                & $<6.0$ \\	
H~1-42   & $8.55^{+0.05}_{-0.03}$ &  $5.97\pm0.25$           & $5.34^{+0.19}_{-0.16}$ & 5.3--6.0 \\	
H~1-50   & $8.69\pm0.04$          &  $5.60^{+0.13}_{-0.16}$  & $5.13^{+0.10}_{-0.12}$ & 5.1--5.6 \\		
Hu~1-1  & $8.61\pm0.03$          & $4.79^{+0.16}_{-0.23}$   & $4.68^{+0.15}_{-0.22}$ & 4.5--4.8 \\	
Hu 2-1   & $8.26^{+0.10}_{-0.02}$ &  $5.57^{+0.05}_{-0.23}$  & $5.29^{+0.04}_{-0.10}$ & 5.3--5.6 \\
IC~418	 & $8.52^{+0.05}_{-0.08}$ & $4.36^{+0.08}_{-0.05}$   & $4.56^{+0.07}_{-0.04}$ & 4.1--4.6 \\
IC~1747	 & $8.62\pm0.04$          &  $<5.72$                 & $<5.06$ 		      & $<5.7$ \\	
IC 2165  & $8.42\pm0.04$          &  $6.05^{+0.06}_{-0.07}$  & $5.49^{+0.06}_{-0.06}$ & 5.5--6.0 \\
IC~3568	 & $8.37\pm0.05$          &  $6.32^{+0.17}_{-0.24}$  & $5.07^{+0.16}_{-0.15}$ & 5.1--6.3 \\	
IC~4191  & $8.72\pm0.04$          &  $5.45\pm0.14$           & $4.97^{+0.11}_{-0.10}$ & 5.0--5.4 \\
IC~4406  & $8.81^{+0.04}_{-0.03}$ &  $<5.04$	             & $<4.97$ 		      & $<5.0$ \\
IC~4593  & $8.57^{+0.06}_{-0.05}$ &  $6.43^{+0.22}_{-0.24}$  & $5.94^{+0.18}_{-0.14}$ & 5.9--6.4 \\
IC~4699  & $8.49^{+0.04}_{-0.03}$ &  $6.02^{+0.21}_{-0.35}$  & $4.98^{+0.24}_{-0.21}$ & 5.0--6.0 \\
IC~4846  & $8.50^{+0.07}_{-0.05}$ &  $5.85^{+0.33}_{-0.39}$  & $5.21^{+0.26}_{-0.24}$ & 5.2--5.8 \\  
IC~5217  & $8.69^{+0.08}_{-0.06}$ &  $6.50^{+0.35}_{-0.51}$  & $5.57^{+0.28}_{-0.36}$ & 5.6--6.5 \\ 
JnEr~1   & $8.54^{+0.18}_{-0.15}$ &  $<5.63$		     & $<5.63$ 		      & $<5.6$ \\
M~1-20   & $8.57^{+0.05}_{-0.04}$ &  $5.36^{+0.22}_{-0.25}$  & $4.92^{+0.17}_{-0.19}$ & 4.9--5.4 \\	
M 1-42   & $8.45^{+0.04}_{-0.03}$ &  $<6.28$                 & $<5.99$                & $<6.3$ \\
M~1-73   & $8.68\pm0.06$          &  $5.88\pm0.09$           & $5.78^{+0.07}_{-0.06}$ & 5.6--5.9 \\
M~2-4    & $8.72^{+0.05}_{-0.04}$ &  $6.27^{+0.13}_{-0.15}$  & $5.94^{+0.08}_{-0.08}$ & 5.9--6.3 \\	
M~2-6    & $8.45^{+0.09}_{-0.05}$ &  $5.99\pm0.17$           & $5.74^{+0.11}_{-0.09}$ & 5.7--6.0 \\	
M~2-27   & $8.87^{+0.05}_{-0.04}$ &  $6.42\pm0.13$           & $6.02^{+0.10}_{-0.08}$ & 6.0--6.4 \\	
M~2-31   & $8.64\pm0.04$          &  $6.74^{+0.16}_{-0.20}$  & $6.16^{+0.14}_{-0.16}$ & 6.2--6.7 \\	
M~2-33   & $8.69^{+0.07}_{-0.03}$ &  $6.50^{+0.27}_{-0.37}$  & $5.92^{+0.23}_{-0.22}$ & 5.9--6.5 \\	
M 2-36   & $8.75\pm0.04$          &  $5.34^{+0.12}_{-0.17}$  & $4.90^{+0.12}_{-0.15}$ & 4.9--5.3 \\
M~2-42   & $8.72\pm0.04$          &  $6.37^{+0.14}_{-0.16}$  & $5.79^{+0.11}_{-0.11}$ & 5.8--6.4 \\	
M~3-7    & $8.69^{+0.05}_{-0.04}$ &  $6.50^{+0.13}_{-0.12}$  & $6.19^{+0.09}_{-0.08}$ & 6.2--6.5 \\	
M~3-29   & $8.48^{+0.06}_{-0.04}$ &  $<6.01$                 & $<5.73$                & $<6.0$ \\	
M~3-32   & $8.60\pm0.04$          &  $<6.99$                 & $<5.89$                & $<7.0$ \\	
MyCn~18  & $8.58\pm0.03$          &  $6.17\pm0.09$           & $5.89^{+0.06}_{-0.05}$ & 5.9--6.2 \\
NGC~40   & $8.62^{+0.06}_{-0.05}$ &  $5.64^{+0.05}_{-0.04}$  & $5.64\pm0.04$          & 5.3--5.6 \\
NGC~2392 & $8.30^{+0.12}_{-0.07}$ &  $6.42^{+0.25}_{-0.17}$  & $6.17^{+0.16}_{-0.12}$ & 6.1--6.4 \\	
NGC~3132 & $8.77^{+0.04}_{-0.03}$ &  $5.51\pm0.08$           & $5.54\pm0.08$          & 5.2--5.5 \\
NGC~3242 & $8.52\pm0.04$          &  $5.87^{+0.10}_{-0.11}$  & $4.99^{+0.08}_{-0.09}$ & 5.0--5.9 \\	
NGC~3587 & $8.53^{+0.13}_{-0.08}$ &  $6.54^{+0.38}_{-0.43}$  & $6.50^{+0.39}_{-0.35}$ & 6.2--6.5 \\
NGC 3918 & $8.68\pm0.04$          &  $5.15^{+0.08}_{-0.10}$  & $4.88^{+0.09}_{-0.10}$ & 4.9--5.2 \\
NGC~5882 & $8.66\pm0.04$          &  $6.31^{+0.10}_{-0.09}$  & $5.53^{+0.07}_{-0.08}$ & 5.5--6.3 \\ 
NGC~6153 & $8.65\pm0.04$          &  $5.90\pm0.11$           & $5.25\pm0.09$          & 5.2--5.9 \\
NGC~6210 & $8.56\pm0.05$          &  $5.80^{+0.13}_{-0.13}$  & $5.25^{+0.09}_{-0.08}$ & 5.2--5.8 \\
NGC~6439 & $8.70\pm0.04$          &  $5.82\pm0.12$           & $5.48^{+0.09}_{-0.10}$ & 5.5--5.8 \\
NGC~6543 & $8.76\pm0.06$          &  $6.27^{+0.15}_{-0.16}$  & $5.61^{+0.10}_{-0.10}$ & 5.6--6.3 \\  
NGC~6565 & $8.74\pm0.03$          &  $6.01\pm0.07$           & $5.84\pm0.06$          & 5.7--6.0 \\
NGC~6572 & $8.59^{+0.06}_{-0.04}$ &  $5.57^{+0.11}_{-0.18}$  & $5.09^{+0.08}_{-0.11}$ & 5.1--5.6 \\  
NGC~6620 & $8.88\pm0.03$          &  $5.65^{+0.13}_{-0.17}$  & $5.49^{+0.12}_{-0.16}$ & 5.4--5.6 \\
NGC~6720 & $8.74^{+0.04}_{-0.03}$ &  $5.21\pm0.08$           & $5.18^{+0.07}_{-0.06}$ & 4.9--5.2 \\   
NGC 6741 & $8.72^{+0.09}_{-0.05}$ &  $6.22\pm0.05$           & $6.17^{+0.20}_{-0.04}$ & 5.9--6.2 \\
NGC~6803 & $8.67^{+0.05}_{-0.04}$ &  $6.01^{+0.10}_{-0.09}$  & $5.47^{+0.07}_{-0.06}$ & 5.5--6.0 \\
NGC 6818 & $8.68\pm0.04$          &  $5.87\pm0.06$           & $5.47\pm0.05$          & 5.5--5.9 \\
NGC~6826 & $8.52\pm0.05$          &  $6.05^{+0.13}_{-0.12}$  & $5.38\pm0.09$ 	      & 5.4--6.0 \\
NGC~6884 & $8.63^{+0.05}_{-0.04}$ &  $6.04^{+0.10}_{-0.11}$  & $5.44\pm0.07$          & 5.4--6.0 \\
NGC~7026 & $8.71\pm0.04$          &  $<5.51$		     & $<5.17$                & $<5.5$ \\
NGC 7662 & $8.43^{+0.06}_{-0.05}$ &  $6.35^{+0.06}_{-0.07}$  & $5.46\pm0.05$          & 5.5--6.4 \\
Vy~2-1   & $8.79^{+0.07}_{-0.03}$ &  $5.64^{+0.13}_{-0.17}$  & $5.31^{+0.14}_{-0.08}$ & 5.3--5.6 
\enddata
\tablenotetext{a}{Derived using equation~(\ref{eq:1}).}
\tablenotetext{b}{Derived using equations~(\ref{eq:2}) and (\ref{eq:3}).}\\
\end{deluxetable}

\subsection{Total abundances without an ICF}
\label{withoutICF}

The ionization correction scheme for iron described above was derived for low-excitation
objects. For high-excitation PNe, like IC~2165 (with a high value of He$^{++}$/He$^+$), the
estimated iron abundances are less reliable. Besides, for those objects with low values of
O$^+$/O$^{++}$, like IC~3568, our method does not constrain very well the iron abundance, since
it remains uncertain by more than an order of magnitude. However, this is the only method that
can be applied at the moment to study in a homogeneous way the iron abundances in a large
sample of PNe. The main reason is that it is difficult to obtain reliable line intensities for
higher ionization states of iron, because they are weak and prone to be blended. 

Some of the nebulae do have detections of lines from iron ionization states higher than
Fe$^{++}$, although there are problems with blends and misidentifications. Nevertheless, 
four of the sample PNe, seem to have reliable measurements of lines from most of their
iron ionization states. For these nebulae, we have derived the iron abundances implied by
the sum of the ionic abundances and compared them to those we obtain using the ICFs described
in \S~\ref{sec:tot_ab}.

The abundance of Fe$^{+3}$ is calculated by solving the equations of statistical equilibrium
for 33 energy levels using the transition probabilities of \citet{FroeseFischer_08} and
the collision strengths of \citet{Zhang_97}. For Fe$^{+4}$ we use a 34-level atom with
transition probabilities from \citet{Nahar_00} and collision strengths from
\citet{Ballance_07}. For Fe$^{+5}$ we use a 19-level atom with transition probabilities and
collision strengths from \citet{Chen_99, Chen_00}. Finally, for Fe$^{+6}$ we use a 9-level
atom with transition probabilities and collision strengths from \citet{Witthoeft_08}. 
Energy levels are those listed in the NIST Atomic Spectra Database\footnote{
http://www.nist.gov/pml/data/asd.cfm};
the H$\beta$ emissivities are based on the empirical formula by \citet{Aller_84}.
We use the values of \dens\ and \temp\oiii\ listed in Table~\ref{tab:1} for all these
calculations.
In what follows, we use air wavelengths from either the NIST database or from the Atomic
Line List\footnote{http://www.pa.uky.edu/$\sim$peter/atomic/} of Peter van Hoof.

NGC~6210, which is a low-excitation PNe with
$I(\mbox{\ion{He}{2}~}\lambda4686)/I(\mbox{H}\beta)=0.015$,
has measurements of [\ion{Fe}{4}]~$\lambda6739.8$, [\ion{Fe}{5}]~$\lambda4227.2$, and
[\ion{Fe}{7}]~$\lambda5276.4$. We think that the last line is a misidentification, since
(1) Fe$^{+4}$ is ionized at 75 eV and this is a low-excitation object, and (2)
\citet{Liu_04b} identify the same feature in NGC~6741, where it leads to an abundance 13
times larger than the one we get from [\ion{Fe}{7}]~$\lambda4893.4$.

Besides [\ion{Fe}{7}]~$\lambda4893.4$, NGC~6741
(with $I(\mbox{\ion{He}{2}~}\lambda4686)/I(\mbox{H}\beta)=0.36$) has measurements of
[\ion{Fe}{5}]~$\lambda4227.2$, and several [\ion{Fe}{6}] lines at 4967.1, 4972.5, 5335.2,
5424.2, 5426.6, 5484.8, 5631.1, and 5677.0 \AA. NGC~6741 also has upper limits for the
intensities of the [\ion{Fe}{4}] lines at 4906.6 and 5233.8 \AA, which imply
$\mbox{Fe}^{+3}/\mbox{H}^+<6\times10^{-7}$, in agreement with the value we estimate by
averaging the abundances of Fe$^{++}$ and Fe$^{4+}$:
$\mbox{Fe}^{+3}/\mbox{H}^+\simeq3.8\times10^{-7}$. We suspect that the two [\ion{Fe}{5}]
lines identified at 3839.3 and 3895.2 \AA\ can be affected by blends or misidentifications
because they lead to high abundances in this nebula and in NGC~6884 and IC~2165.
On the other hand, note that Fe$^{6+}$ has an ionization potential of 125 eV and we do
not expect it to be significantly ionized in any of our PNe.

NGC~6884 (with $I(\mbox{\ion{He}{2}~}\lambda4686)/I(\mbox{H}\beta)=0.18$) has measurements of
[\ion{Fe}{4}]~$\lambda\lambda4900.0,6739.8$, [\ion{Fe}{6}] lines at 5335.2, 5484.8, and
5631.1 \AA, and [\ion{Fe}{7}]~$\lambda4942.5$. There is also an upper limit for the intensity
of [\ion{Fe}{5}]~$\lambda4227.2$, which implies $\mbox{Fe}^{+4}/\mbox{H}^+<1.8\times10^{-7}$,
in agreement with the average of the Fe$^{3+}$ and Fe$^{5+}$ abundances that we adopt:
$\mbox{Fe}^{+4}/\mbox{H}^+\simeq1.1\times10^{-7}$.

Finally, IC~2165, our highest excitation object
($I(\mbox{\ion{He}{2}~}\lambda4686)/I(\mbox{H}\beta)=0.63$), has measurements of
[\ion{Fe}{5}]~$\lambda4227.2$, [\ion{Fe}{6}] lines at 4972.5, 5176.0, 5335.2, 5424.2, 5426.6,
5484.8, 5631.1, and 5677.0 \AA, and [\ion{Fe}{7}]~$\lambda\lambda4988.6,5720.7$. The lines
identified as [\ion{Fe}{6}]~$\lambda4969.0$, [\ion{Fe}{6}]~$\lambda5277.8$, and
[\ion{Fe}{7}]~$\lambda4984.6$ are likely blends or misidentifications. The Fe$^{3+}$
abundance is estimated from the average of the Fe$^{++}$ and Fe$^{4+}$ abundances.
The results obtained for NGC~6741 and NGC~6884 suggest that in this case this is a
reasonable approximation.

We used all the lines listed above which are not affected or suspected of blends and
misidentifications to derive ionic abundances in these four PNe. The resulting ionic and
total abundances are presented in Table~\ref{tab:3b}. If we compare
the iron abundances in this table with those shown in Table~\ref{tab:3}, we find that in
NGC~6741 all the derived values of Fe/H are very similar
($12+\log(\mbox{Fe}/\mbox{H})_{\rm sum}=6.15$
versus $5.9\le12+\log(\mbox{Fe}/\mbox{H})_{\rm ICF}\le6.2$ in Table~\ref{tab:3}),
in NGC~6884 the value of Table~\ref{tab:3b} is intermediate between those of Table~\ref{tab:3}
($12+\log(\mbox{Fe}/\mbox{H})_{\rm sum}=5.65$ versus
$5.4\le12+\log(\mbox{Fe}/\mbox{H})_{\rm ICF}\le6.0$),
in NGC~6210 the iron abundance of Table~\ref{tab:3b} is similar to the upper limit to the iron
abundance inferred from Table~\ref{tab:3} ($12+\log(\mbox{Fe}/\mbox{H})_{\rm sum}=5.80$ versus
$5.2\le12+\log(\mbox{Fe}/\mbox{H})_{\rm ICF}\le5.8$), and in IC~2165 it is close to the
lower limit ($12+\log(\mbox{Fe}/\mbox{H})_{\rm sum}=5.28$ versus
$5.5\le12+\log(\mbox{Fe}/\mbox{H})_{\rm ICF}\le6.1$).
This suggests that the process we follow to constrain the iron abundances within the ranges
shown in the last column of Table~\ref{tab:3} is working well.
The fact that we do not find one ICF working significantly better than the other
might arise from the inherent limitations of all ionization correction schemes, or of this
scheme in particular. On the other hand, further measurements of the main iron ionization
states in other PNe might help in improving this ionization correction scheme or in defining
a better one. In what follows we will check that all our results hold when using any of
the iron ICFs that we have applied.

\begin{deluxetable}{llllll}
\tabletypesize{\footnotesize}
\tablecaption{Iron abundances without ICFs: \{X\} = 12 + $\log$ (X/H)\tablenotemark{a}\label{tab:3b}}
\tablewidth{0pt}
\tablehead{
\colhead{Object} & \colhead{\{Fe$^{+3}$\}} & \colhead{\{Fe$^{+4}$\}} & \colhead{\{Fe$^{+5}$\}}
& \colhead{\{Fe$^{+6}$\}} & \colhead{\{Fe\}}}
\startdata
\objectname{IC~2165}  & 4.62 & 4.66 & 4.58 & 4.35 & 5.28 \\
\objectname{NGC~6210} & 5.75 & 4.04 & -- & -- & 5.80 \\
\objectname{NGC~6741} & 5.58 & 5.45 & 5.10 & 4.67 & 6.15 \\
\objectname{NGC~6884} & 5.21 & 5.04 & 4.77 & 4.67 & 5.65 
\enddata
\tablenotetext{a}{The ionic abundances of Fe$^+$ and Fe$^{++}$ are taken from
Table~\ref{tab:2}.}
\end{deluxetable}

\subsection{Iron depletions}

\begin{figure}
\begin{centering}
\includegraphics[width=10cm,trim = 30 0 -50 0,clip =yes]{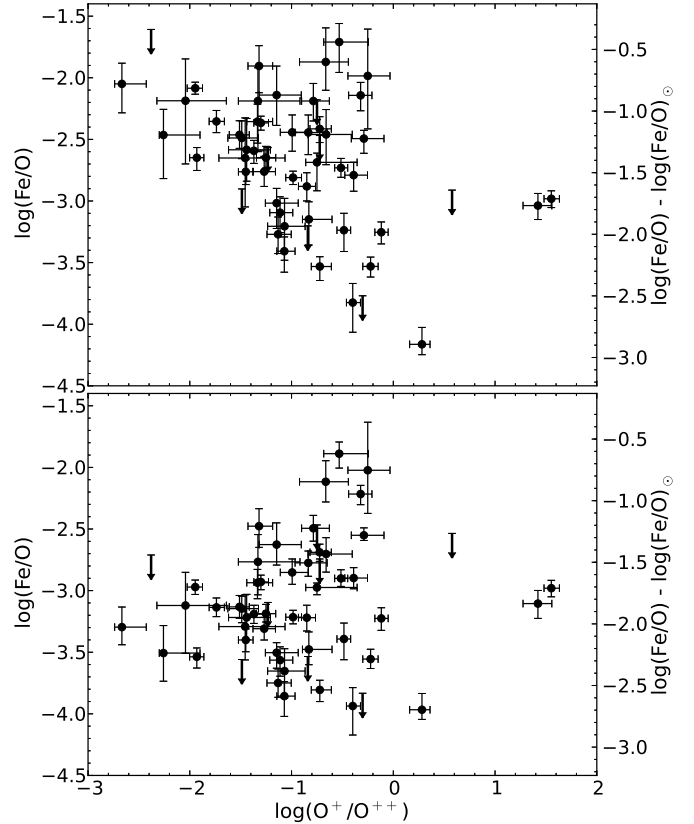}
\vspace{-0.7cm}
\caption{{\small Values of Fe/O (left axis) and the depletion factors for Fe/O 
($[\mbox{Fe/O}]= \log(\mbox{Fe/O}) - \log(\mbox{Fe/O})_{\odot}$, right axis) as a function of 
the degree of ionization for all the PNe. The upper panel shows the values obtained from 
equation~(\ref{eq:1}) whereas the bottom panel shows the ones obtained with equations
(\ref{eq:2}) or (\ref{eq:3}).}
\label{fig:1}}
\end{centering}
\end{figure}

Figure~\ref{fig:1} shows the values of Fe/O  for all the PNe as a function of their degree
of ionization, given by $\log$(O$^{+}$/O$^{++}$).
The upper panel displays the values derived from equation~(\ref{eq:1}) and 
the bottom panel shows the values computed with equations (\ref{eq:2}) and (\ref{eq:3}).
Note that the main difference between the results implied by the two ICFs lies in the iron
abundances of those objects with the lowest values of O$^{+}$/O$^{++}$.

The right axis in Figure~\ref{fig:1} shows estimates of the depletion factors for Fe/O. The
depletion of one particular element (X) is usually calculated as the difference  between
the observed and the expected values of \mbox{X/H}: 
$[\mbox{X}/\mbox{H}]=\log(\mbox{X/H})-\log(\mbox{X/H})_{\rm ref}$.  However, we will use
Fe/O instead of Fe/H to calculate depletion factors, since the intrinsic value of Fe/O is
expected to show less variations from object to object than either Fe/H or O/H
\citep{Ramirez_07}. In principle, we could use a different reference element, but
oxygen abundances require small ionization correction factors in most objects and hence are
the ones that can be determined in a more reliable way. We adopted the solar value of
$\log(\mbox{Fe/O})_{\odot}=-1.27\pm0.11$ \citep{Lodders_10} as the expected abundance ratio
for our objects. For the range of metallicities covered by our sample PNe, the stellar
values of Fe/O are found to decrease from solar to 0.4--0.5 dex below solar in the Galaxy
\citep{Melendez_06, Ramirez_07}.  This means that the halo PN DdDm~1, that has the lowest
metallicity and the highest Fe/O abundance ratio in our sample, could have a depletion
factor close or equal to zero (but its infrared spectrum shows the presence of silicates,
see \S~\ref{sec:dust}). 

We should also bear in mind that the oxygen atoms may be depleted into dust grains, and in such
case our values of $[\mbox{Fe}/\mbox{O}]$ should be lowered by up to $\sim$0.15 dex if oxygen is
trapped in oxides and silicates \citep{Whittet_10}. On the other hand, in \citet{Rodriguez_11}
we found that the oxygen abundances in a group of PNe from the solar neighborhood were
systematically higher than the ones in  nearby \ion{H}{2} regions (calculated either from
collisionally excited lines or recombination lines). We suggested that the difference could be
due to oxygen depletion in organic refractory dust components, but another possible explanation
for these overabundances is oxygen production in the PN progenitor stars, and if this
production is important, it will change significantly the value of Fe/O. However, the amount of
oxygen production by AGB stars is very uncertain and could be negligible
\citep[see the predictions of models by different authors in][]{Karakas_07}. Thus, we present
our results for Fe/O, but in what follows we will check that our results hold both for the Fe/O
and Fe/H abundance ratios.

As we mentioned above, our ionization correction scheme provides three values for the iron 
abundance, and the real value of Fe/O for each object is expected to lie in the range defined
by these three values.  
We can see from Figure~\ref{fig:1} that iron abundances are better constrained for 
objects with a relatively low degree of ionization,
$\log(\mbox{O}^+/\mbox{O}^{++})\gtrsim-1.0$, where the three values obtained with the
different ICFs differ by less than 0.5 dex. The values of $\log(\mbox{Fe/O})$ in our
sample range from $-4.5$ to $-1.7$, which are the most extreme upper and lower limits that
we find.  In the same way, the depletion factors [Fe/O] range from $-3.2$ for IC~418
to $-0.4$ for DdDm~1. 

Even taking into account all the considerations mentioned above that could change the iron
depletions factors shown in the right axes of Figure~\ref{fig:1}, we can conclude that a
significant fraction of our sample PNe have more than $\sim90\%$ of their iron atoms
deposited into dust grains. In agreement with our previous findings in a smaller sample of PNe
\citep{DelgadoInglada_09}, the range of depletions is high, with differences reaching a
factor of $\sim100$.  These differences can be related to the PNe ages or grain compositions,
maybe reflecting different efficiencies of the grain formation and
destruction processes, an issue that we will explore further in the following sections.

\section{The C/O abundance ratios}
\label{sec:co}

As we mentioned in \S~\ref{sec:intro}, one can expect a correlation between the C/O
abundance ratios and the type of dust grains present in PNe. Unfortunately, obtaining
accurate values of C/O for PNe is not easy \citep[see, e.g.,][]{Rola_94, Henry_96}. 
The ionic abundances of carbon and oxygen can be derived from collisionally excited
lines (CELs) or from recombination lines (RLs), and the abundances obtained with RLs
are systematically higher than the ones derived from CELs, by factors that are around
two for many PNe but can reach $\sim$70 \citep[see][]{Liu_06}. 
The reason for this discrepancy is still a matter of debate and we do not 
know which lines lead to more representative abundances in PNe. 
The abundances derived from CELs are highly dependent on the electron 
temperature, but these lines are brighter than RLs and thus, more easily measured. 
On the other hand, RLs are weakly dependent on physical conditions, but 
they are faint and may suffer from other problems 
\citep[see, e.g, ][]{Rodriguez_10,Escalante_12}. 

One important source of uncertainties in the C/O values derived from CELs is 
the normalization between ultraviolet and optical fluxes since carbon lines are 
found in the ultraviolet range whereas oxygen lines are better measured 
in the optical range. This correction introduces uncertainties in the values of C/O, 
which are more severe for extended objects, where ultraviolet and optical observations
may cover different regions of the nebula. One advantage of using RLs is that this
issue does not arise, since they are observed in the optical range.

A further complication is introduced by the ICFs needed to account for unobserved 
ions and estimate the total abundances. For oxygen we have used here the equation 
proposed by \citet{Peimbert_71} to correct for the contribution of O$^{+3}$ to 
the total abundance. In the case of oxygen abundances based on RLs, only the
O$^{++}$ abundances can be easily calculated, and we assume that the distribution
of all ionization states is equal to the one inferred from CELs.
As for carbon, one can apply the widely used correction scheme of \citet{KB_94},
consisting of several ICFs which are employed depending on the degree of 
ionization of the object and the detected ionization stages (up to four different 
ions can be observed in PNe: C$^{+}$, C$^{++}$, C$^{+3}$, and C$^{+4}$), 
although certain cases are not covered by this scheme. Therefore, the correction
scheme of \citet{KB_94} produces an inhomogeneous determination of carbon abundance
when applied to different objects. 

The other method that is frequently adopted to calculate C/O is
$\mbox{C}/\mbox{O}=\mbox{C}^{++}/\mbox{O}^{++}$. However, this ICF generally overestimates
the value of C/O, especially in low ionization PNe \citep{DelgadoInglada_14}. 
Here we calculate C/O in a homogeneous way using the ICF derived in
\citet{DelgadoInglada_14}:
\begin{equation}
\frac{{\rm C}}{{\rm O}} = \frac{{\rm C}^{++}}{{\rm O}^{++}}(0.05 + 2.21\omega -
2.77\omega^2 + 1.74\omega^3),
\label{icf_c}
\end{equation}
where $\omega=\mbox{O}^{++}/(\mbox{O}^{+}+\mbox{O}^{++})$. This ICF is valid in the range
$0.05<\omega<0.97$. In this range, the ICF is expected to reproduce the values of C/O to
within $^{+0.13}_{-0.09}$ dex. However, the C/O values derived for NGC~40 (with
$\omega=0.03$), and for IC~3568, IC~5217, NGC~3242, NGC~5882, M3-32, NGC~7662 (all of them
with $\omega>0.97$) are more uncertain. For NGC~40, we estimate a confidence interval of
$^{+0.26}_{-1.0}$ dex; for the objects with the highest values of $\omega$, we estimate
confidence intervals of $^{+0.26}_{-0.22}$ dex. These uncertainties will be taken into
account in the forthcoming analysis.

The value of the ICF in equation~(5) is close to 1 for most of our objects, where 
$\omega\gtrsim0.8$, changing the C$^{++}$/O$^{++}$ ratio by less than 0.05 dex. However,
in the low ionization PNe IC~418 and NGC~40, the differences between both estimates
reach 0.9 dex.

We compared the values of C/O implied by the three different methods described above
($\mbox{C}/\mbox{O}=\mbox{C}^{++}/\mbox{O}^{++}$, the set of ICFs from \citealt{KB_94}, and
the ICF from \citealt{DelgadoInglada_14}) and by the two types of lines we use (CELs or
RLs). We find that the differences seem to be more related to whether we use CELs or RLs
than to the correction scheme we apply.  Hence we will restrict the forthcoming discussion
to the values of C/O calculated with equation~(5), derived either from  RLs or from
CELs.

\subsection{C/O from CELs}
\label{subsec1:co}

The C$^{++}$ abundances were calculated from the fluxes of the \ion{C}{3}] $\lambda1908$
doublet (provided in the same papers we use for the optical fluxes, see Table~\ref{tab:1})
through the {\it IRAF} routine {\it ionic} using the values of \dens\ and \temp\oiii\ listed
in Table~\ref{tab:1}. 
Since the collision strengths for C$^{++}$, derived by \citet{Berrington_85}, 
are available only for \temp~$>12600$ K, we extrapolated them to lower temperatures 
in order to cover the \temp\ range found for our sample PNe. Table~\ref{tab:4} shows 
the values we obtain for C$^{++}$/H$^{+}$.

The highest differences between our values of $\rm{C}^{++}/\rm{H}^{+}$ and the ones in 
the papers listed in Table~\ref{tab:1} are found for Hu~1-1 (0.35 dex) and IC~4406 (0.23 dex).
For the other PNe the differences are lower than 0.15 dex. These disagreements are due to the 
different physical conditions adopted in the calculations, except for IC~4406, for which the
value given in \citet{Tsamis_03} is a typo, and their corrected 
value agrees with the one we derive (Y.\ Tsamis, private communication).

Using the values of O$^{++}$ from Table~\ref{tab:2}, we calculate the C/O values listed
in the third column of Table~\ref{tab:4}. Because these values are based on CELs located
in the optical ([\ion{O}{3}] $\lambda\lambda4959,5007$) and in the ultraviolet
(\ion{C}{3}] $\lambda1908$), there are several important sources of uncertainty associated
to them, namely, the adopted \temp, the extinction correction, the normalization between
ultraviolet and optical fluxes, and the aperture correction in extended nebulae. 
The uncertainties associated with the normalization of optical and ultraviolet 
fluxes are difficult to quantify. The other sources of uncertainty are discussed below. 

The correction for interstellar extinction is important, and the extinction law for the
bulge PNe, needed for some of our sample objects, is uncertain
\citep[see, e.g.,][and references therein]{Liu_01, Wang_07}. Even a small error in
the extinction coefficient, $\sigma\{c(\mbox{H}\beta)\}=0.1$, introduces an uncertainty
of $\sim0.1$ dex in the intensity ratio of the \ion{C}{3}] $\lambda$1908 and
[\ion{O}{3}] $\lambda$4959 lines. However, the main effect of an error in the extinction
coefficient will occur through its impact on the derived temperature.
The determination of an accurate value of \temp\ is a critical 
factor in obtaining reliable ionic abundances from CELs, and the emissivities of 
ultraviolet CELs are more dependent on \temp\ than those of optical CELs. 
As an example, consider an error of 500 K, which can arise easily from errors in the
extinction correction, the flux calibration or the line measurements that will not necessarily
appear in the quoted line intensity errors. At \temp~$=10000$ K,
this variation introduces a change of $0.18$ dex in C$^{++}$/H$^{+}$ and a change of
$0.07$ dex  in O$^{++}$/H$^{+}$; the final effect in the C$^{++}$/O$^{++}$ ratio will be
around 0.1 dex. 
 
Aperture corrections are needed in those nebulae that are more extended
than the aperture of the International Ultraviolet Explorer ($\sim10\times20$ arsec$^2$),
or that were observed in the optical with a long slit at a single position (as opposed to
a long slit scanning the whole object). 
The scale factors applied by the different authors in our sample PNe go up to 13, 
with NGC~6720 (angular diameter of 76 arcsec) the PN with the 
highest aperture correction \citep{Liu_04a}. We use here the ultraviolet 
fluxes provided by the different authors, already scaled and normalized by them.
The uncertainties related to this procedure are difficult to estimate, but might
reach a factor of 2--3 for objects like NGC~6720. 

Table~\ref{tab:4} shows the C$^{++}$ abundances and the values of C/O based on CELs
and the ICF of equation~(5). We only present in this table the uncertainties associated
with the final values of C/O. We consider an uncertainty of 0.2 dex in
$(\mbox{C}^{++}/\mbox{O}^{++})_{\rm CELs}$, based on the discussion above (it could be
higher for some objects, such as extended nebulae with important aperture corrections),
which is added quadratically to the uncertainty in the ICF computed from equations~(40) and
(41) in \citet{DelgadoInglada_14}.

\setlength{\tabcolsep}{3pt}
\renewcommand{\arraystretch}{1.2}
\begin{deluxetable}{lcl|ccl}
\tabletypesize{\small}
\tablecaption{Ionic abundances: \{X$^{+i}$\} = 12 + $\log$ (X$^{+i}$/H$^{+}$) and 
$\log$(C/O) values from CELs and RLs \label{tab:4}}
\tablewidth{0pt}
\tablehead{
\multicolumn{1}{l}{PN} & \multicolumn{2}{c}{CELs} & \multicolumn{3}{c}{RLs}\\
\multicolumn{1}{l}{} & \multicolumn{1}{c}{\{C$^{++}$\}} & \multicolumn{1}{c}{$\log$(C/O)} & \multicolumn{1}{c}{\{C$^{++}$\}} 
& \multicolumn{1}{c}{\{O$^{++}$\}} & \multicolumn{1}{c}{$\log$(C/O)}}
\startdata
Cn~1-5   &  9.04     & \hspace{0.23cm}$0.28^{+0.20}_{-0.22}$      & 9.08     & 8.90       & \hspace{0.23cm}$0.09^{+0.05}_{-0.10}$ \\
Cn~3-1   &  $\ldots$ & \hspace{0.23cm}$\ldots$  & 8.05     & $\ldots$   & \hspace{0.23cm}$\ldots$ \\
DdDm~1   &  6.78     & $-1.16^{+0.20}_{-0.22}$      & $\ldots$ & 8.43       & \hspace{0.23cm}$\ldots$\\
H~1-41   &  $\ldots$ & \hspace{0.23cm}$\ldots$  & 8.57     & 9.18       & $-0.56^{+0.09}_{-0.10}$ \\
H~1-42   &  7.61     & $-0.86\pm0.22$      & 8.34     & 8.92       & $-0.52\pm0.10$ \\	
H~1-50   &  7.82     & $-0.76^{+0.21}_{-0.22}$      & 8.63     & 9.05       & $-0.38^{+0.09}_{-0.10}$ \\
Hu~1-1   &  $\ldots$ & \hspace{0.23cm}$\ldots$  & 8.94     & 8.54       & \hspace{0.23cm}$0.33^{+0.05}_{-0.10}$ \\
Hu~2-1   &  8.53     & \hspace{0.23cm}$0.34^{+0.21}_{-0.22}$      & 8.62     & 8.71       & $-0.09^{+0.07}_{-0.10}$ \\	
IC~418	 &  8.35     & \hspace{0.23cm}$0.04^{+0.21}_{-0.22}$      & 8.73     & 8.21       & \hspace{0.23cm}$0.26^{+0.06}_{-0.10}$ \\
IC~1747	 &  9.04     & \hspace{0.23cm}$0.55\pm0.22$      & 9.09     & 8.76       & \hspace{0.23cm}$0.39\pm0.10$ \\	
IC~2165  &  8.28     & \hspace{0.23cm}$0.24\pm0.22$      & 8.53     & 8.63       & $-0.04^{+0.09}_{-0.10}$ \\
IC~3568	 &  8.05     & $-0.22\pm0.22$      & 8.49     & 8.75       & $-0.17\pm0.10$ \\	
IC~4191  &  $\ldots$ & \hspace{0.23cm}$\ldots$  & 8.72     & 9.07       & $-0.30^{+0.09}_{-0.10}$ \\
IC~4406  &  8.57     & $-0.10^{+0.20}_{-0.22}$      & 8.89     & 8.89       & $-0.09^{+0.05}_{-0.10}$ \\
IC~4593  &  $\ldots$ & \hspace{0.23cm}$\ldots$  & 8.63     & 8.65       & \hspace{0.23cm}$0.02^{+0.09}_{-0.10}$\\
IC~4699  &  8.00     & $-0.32\pm0.22$       & 8.72     & 9.15      & $-0.35^{+0.11}_{-0.10}$ \\
IC~4846  &  7.86     & $-0.55\pm0.22$      & 8.16     & 8.76       & $-0.52\pm0.10$ \\  
IC~5217  &  8.26     & $-0.29\pm0.22$      & 8.36     & 8.65       & $-0.20\pm0.10$\\ 
JnEr~1   &  $\ldots$ & \hspace{0.23cm}$\ldots$  & $\ldots$ & $\ldots$   &  \hspace{0.23cm}$\ldots$ \\
M~1-20   &  $\ldots$ & \hspace{0.23cm}$\ldots$  & 8.66     & 8.63       & \hspace{0.23cm}$0.08^{+0.08}_{-0.10}$\\
M~1-42   &  7.92     & $-0.44^{+0.21}_{-0.22}$      & 9.38     & 9.63       & $-0.25^{+0.07}_{-0.10}$ \\
M~1-73   &  $\ldots$ & \hspace{0.23cm}$\ldots$  & 8.73     & 9.00       & $-0.34^{+0.05}_{-0.10}$\\
M~2-4    &  $\ldots$ & \hspace{0.23cm}$\ldots$  & 8.50     & 8.89       & $-0.37^{+0.07}_{-0.10}$ \\
M~2-6    &  $\ldots$ & \hspace{0.23cm}$\ldots$  & 7.90     & 8.62       & $-0.74^{+0.06}_{-0.10}$ \\
M~2-27   &  $\ldots$ & \hspace{0.23cm}$\ldots$  & 8.84     & 9.37       & $-0.50^{+0.08}_{-0.10}$ \\
M~2-31   &  $\ldots$ & \hspace{0.23cm}$\ldots$  & 8.79     & $\ldots$   &  \hspace{0.23cm}$\ldots$ \\
M~2-33   & 8.59      & $-0.02\pm0.22$  & 8.30     & 9.04       & $-0.68^{+0.07}_{-0.10}$ \\ 
M~2-36   &  8.65     & $-0.03^{+0.21}_{-0.22}$      & 9.36     & 9.51       & $-0.11^{+0.08}_{-0.10}$ \\ 
M~2-42   &  $\ldots$ & \hspace{0.23cm}$\ldots$  & $\ldots$ & 9.60       & \hspace{0.23cm}$\ldots$ \\
M~3-7     &  $\ldots$ & \hspace{0.23cm}$\ldots$  & $\ldots$ & 9.42       & \hspace{0.23cm}$\ldots$ \\        
M~3-29   &  $\ldots$ & \hspace{0.23cm}$\ldots$  & 8.51     & 9.25       & $-0.75^{+0.07}_{-0.10}$ \\
M~3-32   &  8.42       & $-0.06\pm0.22$  & 9.55     & 9.74       & $-0.10^{+011}_{-0.10}$ \\
MyCn~18  &  $\ldots$ & \hspace{0.23cm}$\ldots$  & 8.36     & 8.81       & $-0.47^{+0.07}_{-0.10}$ \\
NGC~40   &  8.01     & $-0.02^{+0.24}_{-0.22}$      & 8.81     & 8.61       & $-0.76^{+0.14}_{-0.10}$ \\
NGC~2392 &  7.69     & $-0.39^{+0.21}_{-0.22}$     & $\ldots$ & 9.81       & \hspace{0.23cm}$\ldots$\\
NGC~3132 &  8.18     & $-0.47^{+0.20}_{-0.22}$     & 8.82     & 8.80       & $-0.11^{+0.05}_{-0.10}$\\
NGC~3242 &  8.04     & $-0.28\pm0.22$      & 8.79     & 8.78       & \hspace{0.23cm}$0.09^{+0.09}_{-0.10}$\\
NGC~3587 &  $\ldots$ & \hspace{0.23cm}$\ldots$  & 8.36     & 9.38     & $-1.13^{+0.05}_{-0.10}$ \\
NGC~3918 &  8.36     & $-0.07^{+0.21}_{-0.22}$      & 8.70     & 8.75       & $-0.06^{+0.07}_{-0.10}$ \\
NGC~5882 &  8.11     & $-0.46\pm0.22$      & 8.58     & 8.93       & $-0.27\pm0.10$ \\
NGC~6153 &  8.38     & $-0.16\pm0.22$      & 9.35     & 9.51       & $-0.09\pm0.10$ \\
NGC~6210 &  7.87     & $-0.60\pm0.22$      & 8.80     & 9.53       & $-0.66\pm0.10$ \\
NGC~6439 &  8.32     & $-0.24^{+0.21}_{-0.22}$      & 8.99     & 9.21       & $-0.21^{+0.07}_{-0.10}$ \\
NGC~6543 &  8.48     & $-0.19\pm0.22$      & 8.76     & 9.07       & $-0.24\pm0.10$ \\
NGC~6565 &  8.29     & $-0.32^{+0.20}_{-0.22}$      & 8.65     & 8.85       & $-0.25^{+0.05}_{-0.10}$ \\
NGC~6572 &  8.77     & \hspace{0.23cm}$0.25^{+0.21}_{-0.22}$      & 8.69     & 8.71       & \hspace{0.23cm}$0.03^{+0.09}_{-0.10}$ \\
NGC~6620 &  8.16     & $-0.58^{+0.20}_{-0.22}$      & 8.94     & 9.24       & $-0.36^{+0.05}_{-0.10}$ \\
NGC~6720 &  8.38     & $-0.19^{+0.20}_{-0.22}$      & 8.94     & 8.90       & $-0.08^{+0.05}_{-0.10}$ \\
NGC~6741 &  8.40     & $-0.09^{+0.20}_{-0.22}$      & 8.79     & 8.74       & $-0.04^{+0.05}_{-0.10}$ \\
NGC~6803 &  8.24     & $-0.33\pm0.22$      & 8.79     & 9.05       & $-0.21^{+0.09}_{-0.10}$ \\
NGC~6818 &  8.17     & $-0.14\pm0.22$      & 8.66     & 8.57       & \hspace{0.23cm}$0.12^{+0.08}_{-0.10}$ \\
NGC~6826 &  8.40     & $-0.03\pm0.22$      & 8.73     & 8.92       & $-0.11\pm0.10$ \\
NGC~6884 &  8.45     & $-0.02\pm0.22$      & 8.87     & 8.98       & $-0.05^{+0.09}_{-0.10}$ \\
NGC~7026 &  8.33     & $-0.26^{+0.21}_{-0.22}$      & 8.93     & 9.07       & $-0.13^{+0.07}_{-0.10}$ \\
NGC~7662 &  7.99     & $-0.15^{+0.21}_{-0.22}$      & 8.67     & 8.58       & \hspace{0.23cm}$0.17\pm0.10$ \\
Vy~2-1   &  8.61     & $-0.11^{+0.21}_{-0.22}$      & 8.62     & 8.98       & $-0.34^{+0.07}_{-0.10}$ 
\enddata                                      
\end{deluxetable}                             
\renewcommand{\arraystretch}{1}

\subsection{C/O from RLs}
\label{subsec2:co}

\ion{C}{2} and \ion{O}{2} RLs can be measured in the optical range and do not suffer
from many of the uncertainties discussed above. Therefore, these lines will provide
better estimates of the C/O abundance ratio, or they will do so if they can be considered
to sample the abundances of the nebular gas. For example, these emission lines might not
be representative of the nebular abundances if they arise mostly from metal-rich inclusions,
whose existence in PNe has been postulated to explain the discrepant abundances implied by
CELs and RLs \citep[see, e.g.,][]{Liu_04a}.

The C$^{++}$/H$^{+}$ and O$^{++}$/H$^{+}$ abundance ratios implied by RLs are derived for
all objects with available measurements of the required RLs using the values of \dens\ and
\temp\oiii\ in Table~\ref{tab:2} and the effective recombination coefficients for H$\beta$ of
\citet{Storey_95}. 
The C$^{++}$ abundances are calculated using the \ion{C}{2} $\lambda$4267 line and
the case B effective recombination coefficients of \citet{Davey_00}. 
The O$^{++}$ abundances are computed from the total intensity of multiplet 1 of \ion{O}{2}
and the recombination coefficients of \citet{Storey_94}. The multiplet intensity was
corrected for the contribution of undetected lines with the formulae derived by
\citet{Peimbert_05}. 
Typical errors in C$^{++}$/H$^{+}$ and O$^{++}$/H$^{+}$ arising from errors in the line
intensities are $\pm0.02$ dex and $\pm0.06$ dex, respectively. Final errors in 
C$^{++}$/O$^{++}$ due to errors in the line intensities are around $\pm0.06$ dex. If we
also consider the uncertainties in the ICF, as explained in \S~\ref{subsec1:co}, we
obtain the final uncertainties presented in Table~\ref{tab:4} together with the ionic
abundances of C$^{++}$ and O$^{++}$ obtained from RLs.

We found differences between our values and the ones in the reference papers 
that are in general lower than 0.06 dex but reach 0.17 dex for some 
nebulae. The highest differences are found for H~1-42 (with a difference of 0.32 dex 
in the value of C$^{++}$/H$^{+}$), M~2-42 (0.56 dex in O$^{++}$/H$^{+}$), and M~3-29 
(0.49 dex in O$^{++}$/H$^{+}$) and are not due to differences in the physical conditions
used in the calculations. We do not know the reasons for these discrepancies, but
they do not affect our conclusions, since we do not have information 
about the types of grains present in H~1-42 and M~3-29, and we cannot calculate the
C/O abundance ratio in M~2-42. Hence, these PNe are not included in the forthcoming 
analysis. 

\begin{figure}
\includegraphics[width=9cm,trim = 20 0 20 20,clip =yes]{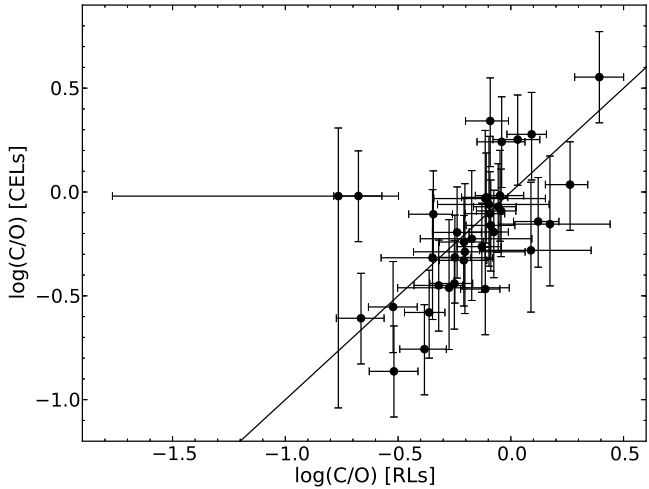}
\caption{{\small Comparison of C/O = C$^{++}$/O$^{++}$ values derived from CELs and from 
RLs. The error bars represent typical errors as explained in the text. The line shows
where the values of C/O are equal. \label{fig:c2o2}}}
\end{figure}

Figure~\ref{fig:c2o2} shows a comparison between the values of C/O obtained from
CELs and RLs in those PNe where both types of lines are observed.
We also plot in the figure the error bars associated with this abundance ratio
for each object.
The error bars take into account both the uncertainties in the line intensity ratios
($\pm0.2$ dex for the values derived from CELs and $\pm0.06$ dex for the ones derived
from RLs) and the uncertainties in the ICF.
Although there is general agreement between these two abundance ratios, a result
previously found in several studies \citep[see][and references therein]{Liu_04a},
the differences are high for some objects.
The highest differences reach $\sim0.7$ dex and are found for NGC~40 and M~2-33. 
NGC~40 is an extended PN (diameter of around 48 arcsec) with a Wolf-Rayet central star,
thus, an incorrect aperture correction and/or contamination of the ultraviolet line
fluxes with stellar winds could explain the discrepancy. M~2-33 is smaller (with a
diameter of around 5.8 arcsec) and thus, we expect that the scaling of optical and ultraviolet
observations is not that critical. The discrepancy could also be
explained if the \temp\ is underestimated by $\sim3000$ K, but we do not expect such a
large error in \temp. We have explored if the differences between the ionic abundance
ratios  derived from CELs and RLs are related to parameters such as \temp\oiii, \dens, 
the extinction coefficient, the nebular size, the type of dust, the abundance
discrepancy factor, or the  \temp\ deduced from the \ion{H}{1} recombination continuum
Balmer discontinuity, but we do not find any obvious correlation. 

As we mentioned above, the C/O abundance ratios in PNe tell us whether the ionized 
gas is carbon rich or oxygen rich. We find that both CELs and RLs imply that 20\% of
the PNe are C--rich. Most of the PNe in our sample seem to be O-rich objects.
According to theoretical models \citep[see, e.g.,][]{Marigo_03}, these PNe evolve either
from the lowest mass progenitors (masses below 1.5 M$_{\odot}$) or from the highest
mass progenitors (above 4--5 M$_{\odot}$), although the details depend on the metallicity
and on the model assumptions.

\section{Dust features from infrared spectra} 
\label{sec:dust}  
                                      
We used the archive of the Infrared Spectrograph (IRS; \citealt{Houck_04}), on board the
{\it Spitzer Space Telescope} \citep{Werner_04}, to download the spectra available for
our sample PNe. We decided to look for the following dust features: polycyclic aromatic
hydrocarbons (PAHs), the broad features around 11 and 30 $\mu$m (usually associated with
SiC and MgS, respectively), amorphous silicates, and crystalline silicates. 
We restricted the analysis to these features because they are relatively easy to identify 
and are, in principle, reliably associated with either a carbon-rich or an 
oxygen-rich environment (we will come back to this point later). A detailed analysis of all 
the dust features present in the PNe is beyond the scope of this paper.
We also compiled dust identifications from the literature, mainly from spectra of the 
{\it Infrared Space Observatory} ({\it ISO\/}; \citealt{Kessler_96}). The results are 
summarized in Table \ref{tab:5} and are discussed below.

\setlength{\tabcolsep}{6pt}
\renewcommand{\arraystretch}{1.2}
\begin{deluxetable*}{lllcccccl}
\tabletypesize{\footnotesize}
\tablecaption{Compilation of dust features\tablenotemark{a}, range of {\rm Fe/O} values: \{{\rm Fe/O}\} = $\log$({\rm Fe/O}), and $\log$(C/O) values. \label{tab:5}}
\tablewidth{0pt}
\tablehead{
\colhead{Object}      & \colhead{$\Delta$\{Fe/O\}} & \colhead{$\log$(C/O)} & \colhead{PAHs} & \colhead{SiC} & \colhead{30 $\mu$m} & \colhead{Amorphous} & \colhead{Crystalline} & \colhead{Source}\\
\colhead{}                & \colhead{}  & \colhead{CELs/RLs} & \colhead{}     & \colhead{}    & \colhead{} & \colhead{silicates}   &\colhead{silicates}  & 
}
\startdata
Cn~1-5    & $[-2.4, -2.1]$ & $+0.28$/$+0.09$ & $\surd$         & $\times$    & $\times$    & $\times$ & $\surd$   & {\it ISO} (1), {\it Spitzer} (2,3), UKIRT (7) \\
DdDm~1   & $[-2.0, -1.7]$ & $-1.16$/$\ldots$ & $\times$        & $\times$    & $\times$    & $\surd$  & $\surd$   & {\it Spitzer} (3,4)                      \\
H~1-50    & $[-3.6, -3.1]$ & $+0.34$/$-0.09$ & $\surd$?        & $\times$    & $\times$    & $\surd$ & $\surd$   & {\it Spitzer} (2,3)  		     \\ 	  
Hu~2-1    & $[-3.0, -2.7]$ & $+0.34$/$-0.09$ & $\surd$         & $\surd$     & $\ldots$    & $\times$ & $\ldots$  & {\it ISO} (1),  {\it Spitzer} (3), UKIRT (6)\\
IC~418    & $[-4.5, -4.0]$ & $+0.04$/$+0.26$  & $\surd$         & $\surd$    & $\surd$     & $\times$ & $\ldots$  & {\it ISO} (1, 5), {\it Spitzer} (3), UKIRT (6)  \\
IC~2165  & $[-2.9, -2.4]$ & $+0.24$/$-0.04$ & $\times$        & $\surd$?     & $\ldots$    & $\times$ & $\ldots$  & {\it Spitzer} (3), UKIRT (6)            \\
IC~3568  & $[-3.3, -2.0]$ & $-0.22$/$-0.17$ & $\times$        & $\ldots$    & $\ldots$    & $\ldots$ & $\ldots$  & {\it ISO} (1)                            \\
IC~4406  & $<-3.8$      & $-0.10$/$-0.09$ & $\times$        & $\ldots$    & $\ldots$    & $\ldots$ & $\ldots$  & {\it ISO} (1)                            \\
IC~4846  & $[-3.3, -2.6]$ & $-0.55$/$-0.52$ & $\times$        & $\times$    & $\times$    & $\surd$? & $\times$? & {\it Spitzer} (3)                        \\  
M~1-20   & $[-3.6, -3.2]$ & $\hspace{0.21cm}\ldots$/$+0.08$ & $\surd$         & $\surd$     & $\surd$     & $\times$ & $\times$  & {\it Spitzer} (2,3,8), UKIRT (7)        \\	
M~1-42   & $<-2.2$        & $-0.44$/$-0.25$ & $\surd$         & $\times$    & $\times$    & $\times$ & $\surd$   & {\it ISO} (1), {\it Spitzer} (3)         \\
M~2-27   & $[-2.8, -2.4]$ & $\hspace{0.21cm}\ldots$/$-0.50$ & $\surd$         & $\times$    & $\times$    & $\times$ & $\surd$   & {\it Spitzer} (2,3)                      \\	
M~2-31   & $[-2.5, -1.9]$ & $\hspace{0.21cm}\ldots$/$\ldots$ & $\surd$         & $\times$    & $\times$    & $\times$ & $\surd$   & {\it Spitzer} (3,9)                      \\	
M~2-36   & $[-3.9, -3.4]$ & $-0.03$/$-0.11$ & $\times$        & $\ldots$    & $\ldots$    & $\ldots$ & $\ldots$  & {\it ISO} (1)                            \\
M~2-42   & $[-2.9, -2.4]$ & $\hspace{0.21cm}\ldots$/$\ldots$ & $\times$        & $\times$    & $\times$    & $\times$ & $\surd$   & {\it Spitzer} (3)                        \\ 
MyCn~18  & $[-2.7, -2.4]$ & $\hspace{0.21cm}\ldots$/$-0.47$ & $\surd$         & $\times$    & $\times$    & $\surd$  & $\surd$   & {\it Spitzer} (3)                        \\
NGC~40\tablenotemark{b}    & $[-3.3, -3.0]$ & $-0.02$/$-0.76$ & $\surd$         & $\times$    & $\surd$     & $\times$ & $\times$  & {\it ISO} (1, 5), {\it Spitzer} (3)      \\
NGC~2392  & $[-2.2, -1.9]$ & $-0.39$/$\ldots$ & $\times$        & $\times$    & $\times$    & $\times$ & $\times$  & {\it Spitzer} (3)                        \\	
NGC~3132  & $[-3.6, -3.2]$ & $-0.47$/$-0.11$ & $\times$        & $\times$    & $\times$    & $\times$ & $\surd$   & {\it Spitzer} (3)                       \\
NGC~3242  & $[-3.5, -2.6]$ & $-0.28$/$+0.09$ & $\times$        & $\times$    & $\surd$     & $\times$ & $\times$  & {\it Spitzer} (3)                      \\
NGC~3918  & $[-3.8, -3.5]$ & $-0.07$/$-0.06$ & $\times$        & $\times$    & $\surd$     & $\times$ & $\ldots$  & {\it ISO} (1, 5, 10), {\it Spitzer} (3) \\
NGC~6153  & $[-3.4, -2.8]$ & $-0.16$/$-0.09$ & $\times$        & $\ldots$    & $\ldots$    & $\surd$? & $\times$? & {\it ISO} (1, 10)                         \\
NGC~6210  & $[-3.3, -2.8]$ & $-0.60$/$-0.66$ & $\times$        & $\times$    & $\times$    & $\times$? & $\surd$   & {\it ISO} (1),  {\it Spitzer} (3)       \\
NGC~6439  & $[-3.2, -2.9]$ & $-0.24$/$-0.21$ & $\surd$?        & $\times$    & $\times$    & $\times$ & $\surd$   &  {\it Spitzer} (3)                      \\ 
NGC~6543  & $[-3.1, -2.5]$ & $-0.19$/$-0.24$ & $\times$        & $\ldots$    & $\ldots$    & $\surd$  & $\surd$   & {\it ISO} (1, 10)                          \\  
NGC~6572  & $[-3.5, -3.0]$ & $+0.25$/$+0.03$ & $\times$        & $\surd$?    & $\ldots$    & $\times$ & $\times$  & {\it ISO} (1), UKIRT (6)                \\  
NGC~6720  & $[-3.8, -3.5]$ & $-0.19$/$-0.08$ & $\times$        & $\times$    & $\ldots$    & $\times$ & $\ldots$  & {\it ISO} (1),  {\it Spitzer} (3)       \\   
NGC~6741  & $[-2.8, -2.5]$ & $-0.09$/$-0.04$ & $\surd$         & $\times$    & $\ldots$    & $\times$ & $\ldots$  & {\it ISO} (1), {\it Spitzer} (3), UKIRT (6)\\
NGC~6818  & $[-3.2, -2.8]$ & $-0.14$/$+0.12$ & $\times$        & $\times$    & $\times$    & $\times$ & $\times$?  & {\it Spitzer} (3)                        \\
NGC~6826  & $[-3.1, -2.5]$ & $-0.03$/$-0.11$ & $\times$        & $\times$    & $\surd$     & $\times$ & $\ldots$  & {\it ISO} (1, 5),  {\it Spitzer} (3)     \\
NGC~6884  & $[-3.2, -2.6]$ & $-0.02$/$-0.05$ & $\surd$         & $\times$    & $\ldots$    & $\times$ & $\ldots$  & {\it ISO} (1),  {\it Spitzer} (3)        \\
NGC~7026  & $<-3.2$ & $-0.26$/$-0.13$ & $\surd$         & $\times$    & $\times$    & $\times$ & $\surd$   &  {\it Spitzer} (3)                       \\
NGC~7662  & $[-3.0, -2.1]$ & $-0.15$/$+0.17$ & $\times$        & $\times$    & $\ldots$    & $\times$ & $\ldots$  & {\it ISO} (1),  {\it Spitzer} (3)                                                   
\enddata
\tablenotetext{a}{$\surd$: positive identification, $\surd$?: possible identification, 
$\times$?: doubtful identification, $\times$: no identification, $\ldots$: no 
spectral information.}
\tablenotetext{b}{The 30 $\mu$m feature is not present in the {\it Spitzer}/IRS spectrum but it was identified by 
\citet{Hony_02} using {\it ISO} spectra (see text).}
\tablerefs{(1) \citet{Cohen_05}, (2) \citet{PC_09}, (3) this work, (4) \citet{Henry_08}, 
(5) \citet{Hony_02}, (6) \citet{Casassus_01a}, (7) \citet{Casassus_01b}, (8) \citet{GH_10}, 
(9) \citet{Gutenkunst_08}, (10) \citet{BS_05}, (11) \citet{Stanghellini_12}}
\end{deluxetable*}
\renewcommand{\arraystretch}{1.}

\subsection{Data from {\it Spitzer}}

We found {\it Spitzer} IRS spectra for 27 PNe of our sample (we excluded the very noisy
spectra of JnEr~1 and NGC~3587).  
The data belong to the following observing programs: ID 45 (PI: T. Roellig), 
ID 93 (PI: D. Cruikshank), ID 1406 (PI: L. Armus),  ID 1427 (calibration program), ID 20049 
(PI: K. Kwitter), ID 30285, 40115 (PI: G. Fazio), ID 30430, 40536 (PI: H. Dinnerstein), 
ID 30550 (PI: J. R. Houck), ID 3633 (PI: M. Bobrowsky), and ID 50261 (PI: L. Stanghellini). 
The {\it Spitzer} spectra of Cn~1-5, DdDm~1, H~1-50, IC~4846, M~1-20, M~2-27, M~2-42, 
M~2-31, and MyCn~18 have been studied already in several works
\citep{PC_09, Gutenkunst_08, Henry_08, Stanghellini_12}, where some of the dust
features in these objects are identified. 

The PNe were observed with at least one of the four different modules of IRS (SL, LL, 
SH, and LH) and, therefore, the wavelength coverage and the spectral resolution are not
the same for all the nebulae. Each module is named by its wavelength coverage and
resolution as Short-Low (SL, covering the range 5.2--14.5 $\mu$m with a spectral resolution 
$\lambda$/$\Delta\lambda$ = 64--128), Long-Low 
(LL: 14.0--38 \micron\ and the same spectral resolution as SL), Short-High (SH: 9.9--19.6
$\mu$m and $\lambda$/$\Delta\lambda$ $\sim$ 600), and Long-High (LH: 18.7--37.2 $\mu$m and 
the same spectral resolution as SH). We used the observations performed in the {\it staring} 
mode in which the object is placed at two different positions (nods) in the slit. 

We retrieved the data from the {\it Spitzer}/IRS archive and we reduced them following 
the usual steps\footnote{We used the packages and tools available at 
http://irsa.ipac.caltech.edu/data/SPITZER/docs/.}.
We started the reduction from the Base Calibrated Data (BCD). 
First, we used the IDL procedure {\tt doall\_coads} to produce a single coadded image 
for each nod position and module separately. Each of the low-resolution modules contains 
two slits; when one of them is observing the source, the other one is off-source.
We used the off-source spectra to subtract the background in the on-source spectra. 
For the high-resolution modules, extra sky images are needed to remove this contribution 
in the source spectra. In some cases, there were no available images to remove the 
sky background, but this is not important for our identification purposes. Rogue pixels were 
removed with the {\tt irsclean} tool. 
The spectra for each nod position was wavelength and flux calibrated, and extracted 
with the {\it Spitzer} IRS Custom Extractor (SPICE) using one of the two available extractions: 
point-source aperture or full-slit extraction, depending on the size of the PNe. 
Finally, we used the Spectroscopic Modeling Analysis and Reduction Tool (SMART, 
\citealt{Higdon_04,Lebouteiller_10}) to manually eliminate bad pixels, jumps, and to 
combine and merge into one final spectra the observations of each module. High-resolution
spectra were smoothed using a box-car algorithm (of widths 0.06 $\mu$m and 0.08 $\mu$m for
the SH and LH modules, respectively) to slightly lower the noise without losing information 
about the dust features.

Different regions of the final {\it Spitzer}/IRS spectra for the 27 sample PNe
are shown in Figures~\ref{fig:dust_2}--\ref{fig:dust_5}. For the convenience of the reader,
we display all the {\it Spitzer} spectra available for our objects, although some
of them have already been published, and some of the features of interest have been
identified in the works we mentioned above. A total of 18 PNe have observations for the
whole IRS spectral range, 5--37 $\mu$m, where many dust features can be observed
(Figs.~\ref{fig:dust_3} and \ref{fig:dust_4}). The other nebulae do not have data
for the full spectral range that can be covered with {\it Spitzer}/IRS. On the other hand, some
objects have wavelength ranges covered at both high and low spectral resolutions. In such
cases, we show the spectra where the dust emission features can be seen more clearly.

\subsection{Identification of dust features}

\subsubsection{C-rich features: SiC, the 30 $\mu$m feature, and PAHs}

\begin{figure}
\begin{center}
\includegraphics[width=8cm,trim = 10 0 0 0,clip =yes]{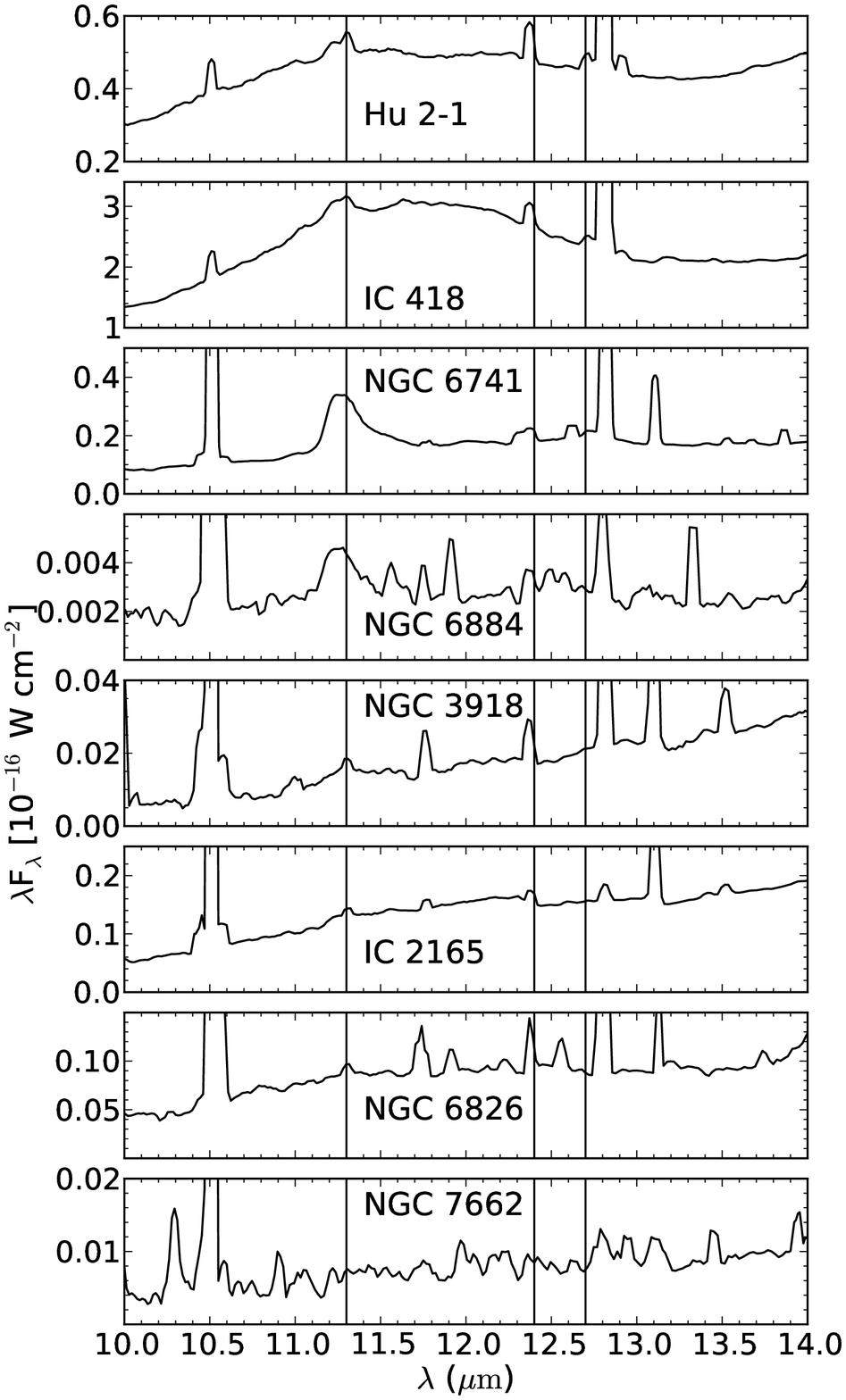}
\caption{{\it Spitzer}/IRS spectra showing PAH and/or SiC features. The solid lines
are at the wavelengths of the usual peak intensities of the PAH features at 11.3, 12.4, and 12.7
$\mu$m. The spectra available for these PNe do not cover the same range as the ones for 
the objects shown in Figures \ref{fig:dust_3} and \ref{fig:dust_4}. \label{fig:dust_2}}
\end{center}
\end{figure}

\begin{figure*}
\begin{center}
\includegraphics[width=16cm,trim = 85 0 0 0,clip =yes]{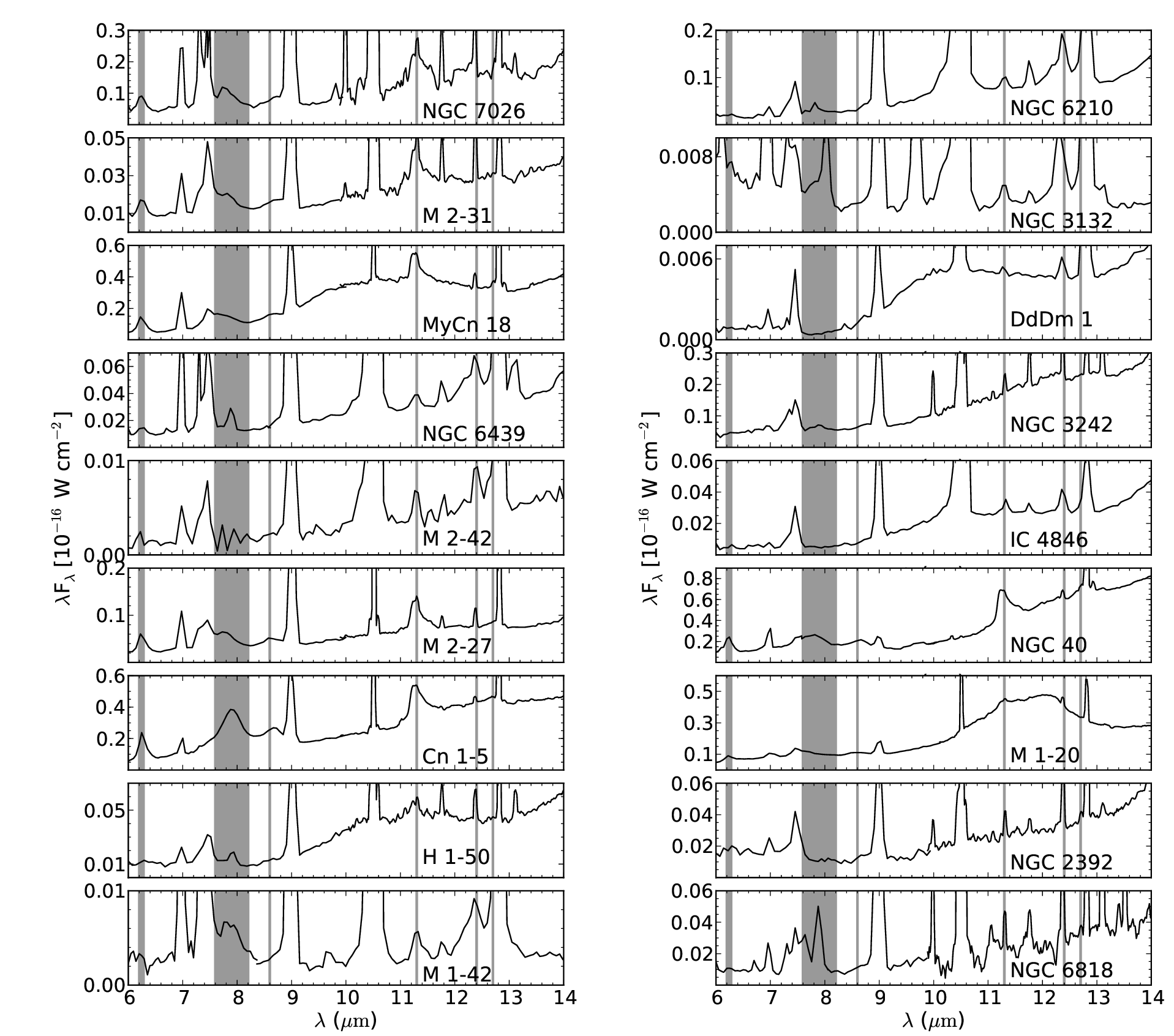}
\caption{{\it Spitzer}/IRS spectra showing PAH features. 
Thin vertical lines represent the usual peak intensity of the features at 8.6, 11.3, 12.4, and
12.7 $\mu$m, whereas shaded areas show the range of variation for peak intensity of the
features at 6.2 and 7.7 $\mu$m. The PNe are ordered as in figure \ref{fig:dust_4}.
\label{fig:dust_3}}
\end{center}
\end{figure*}

Using the IRS spectra, we identify the broad feature around 11.3 \micron, 
associated with SiC grains, in Hu~2-1, IC~418, and M~1-20 (Figs.~\ref{fig:dust_2}--\ref{fig:dust_3}).
This feature was already detected by \citet{Casassus_01a, Casassus_01b} and 
\citet{Cohen_05} in UKIRT and ISO spectra of these three PNe, and also in IC~2165 
and NGC~6572. In the last two PNe, the feature is not as evident as for the other objects.
In the case of IC~2165, there are {\it Spitzer}/IRS spectra, and we do not see clear
evidence of the feature (Fig.~\ref{fig:dust_2}). Therefore, we label the identification
as doubtful in this PNe. 

\begin{figure*}
\begin{center}
\includegraphics[width=16cm,trim = 85 0 0 0,clip =yes]{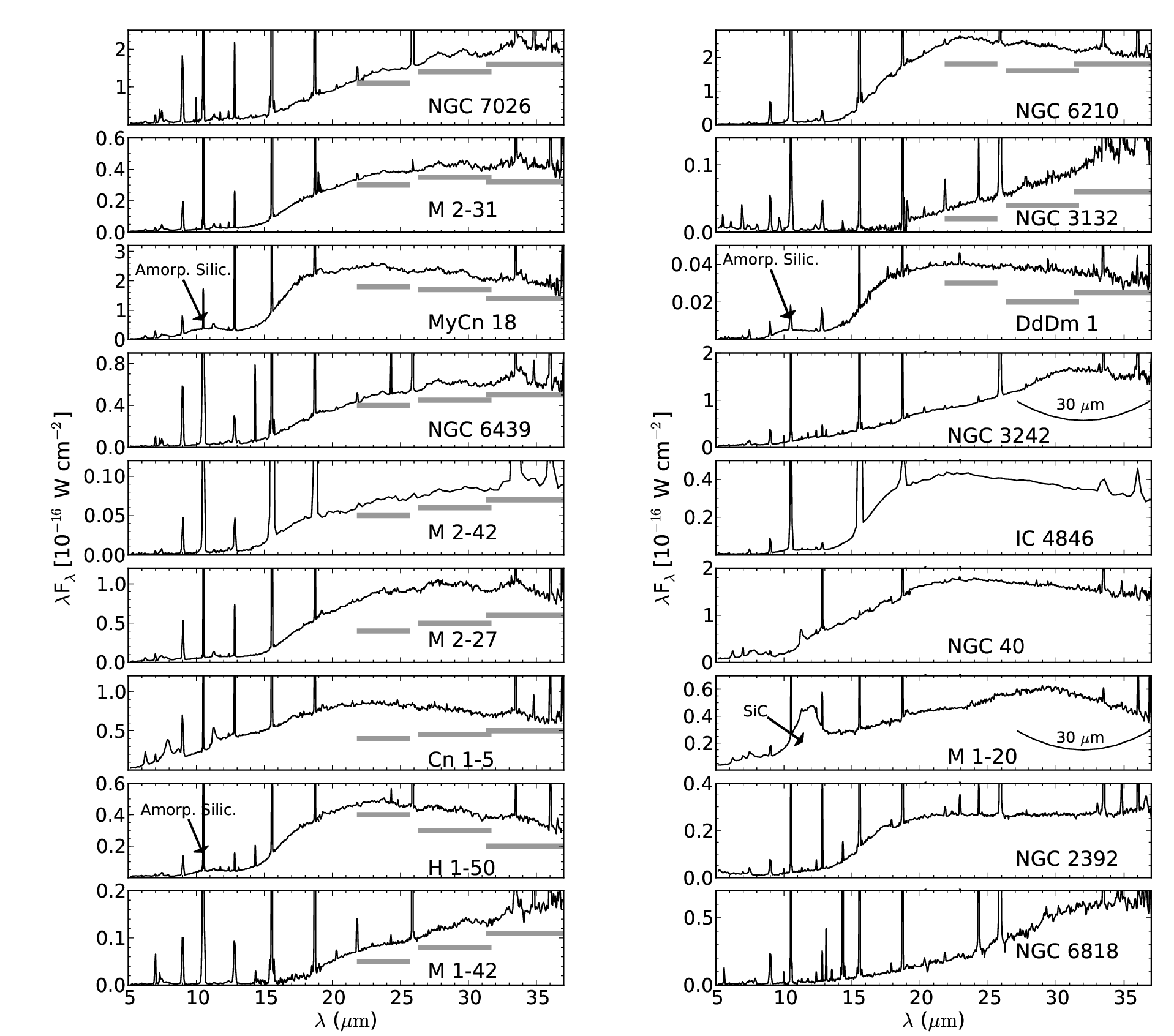}
\caption{{\it Spitzer}/IRS spectra showing crystalline silicate complexes around 23, 28, and
33 $\mu$m (grey rectangles), amorphous silicates, SiC, and the broad feature at 30 $\mu$m. 
The PNe are ordered according to the prominence of their crystalline silicate features.
\label{fig:dust_4}}
\end{center}
\end{figure*}

The 30 $\mu$m feature, often attributed to MgS grains, is present in the spectra of
NGC~3242 and M~1-20 (see Fig.~\ref{fig:dust_4}). Besides, \citet{Hony_02} detected
this feature in {\it ISO} spectra of IC~418, NGC~40, NGC~3918, and NGC~6826. 
We do not observe the feature in NGC~40, but this could be due to the different 
areas observed by {\it ISO} ($\sim$ 440 arcsec$^2$, covering almost the whole nebula) 
and {\it Spitzer} ($\sim$ 248 arcsec$^2$).

There are PAH features in the IRS spectra of 12 of the 27 PNe with available data
(see Figs.~\ref{fig:dust_2}--\ref{fig:dust_3}), and we cannot rule out their presence in
at least 2 of the other PNe. Some of the PNe would require deeper spectra or a higher
spectral resolution to reach more definitive conclusions. As an example, \citet{Cohen_05}
did not find PAHs in M~1-42 using {\it ISO} observations, but we think that the
{\it Spitzer} spectra of this nebula shows evidence of these features
(see Fig.~\ref{fig:dust_3}). For this nebula, a higher spectral resolution would show
the presence or absence of PAHs more clearly. 

\begin{figure}
\includegraphics[width=8cm,trim = 0 0 20 15,clip =yes]{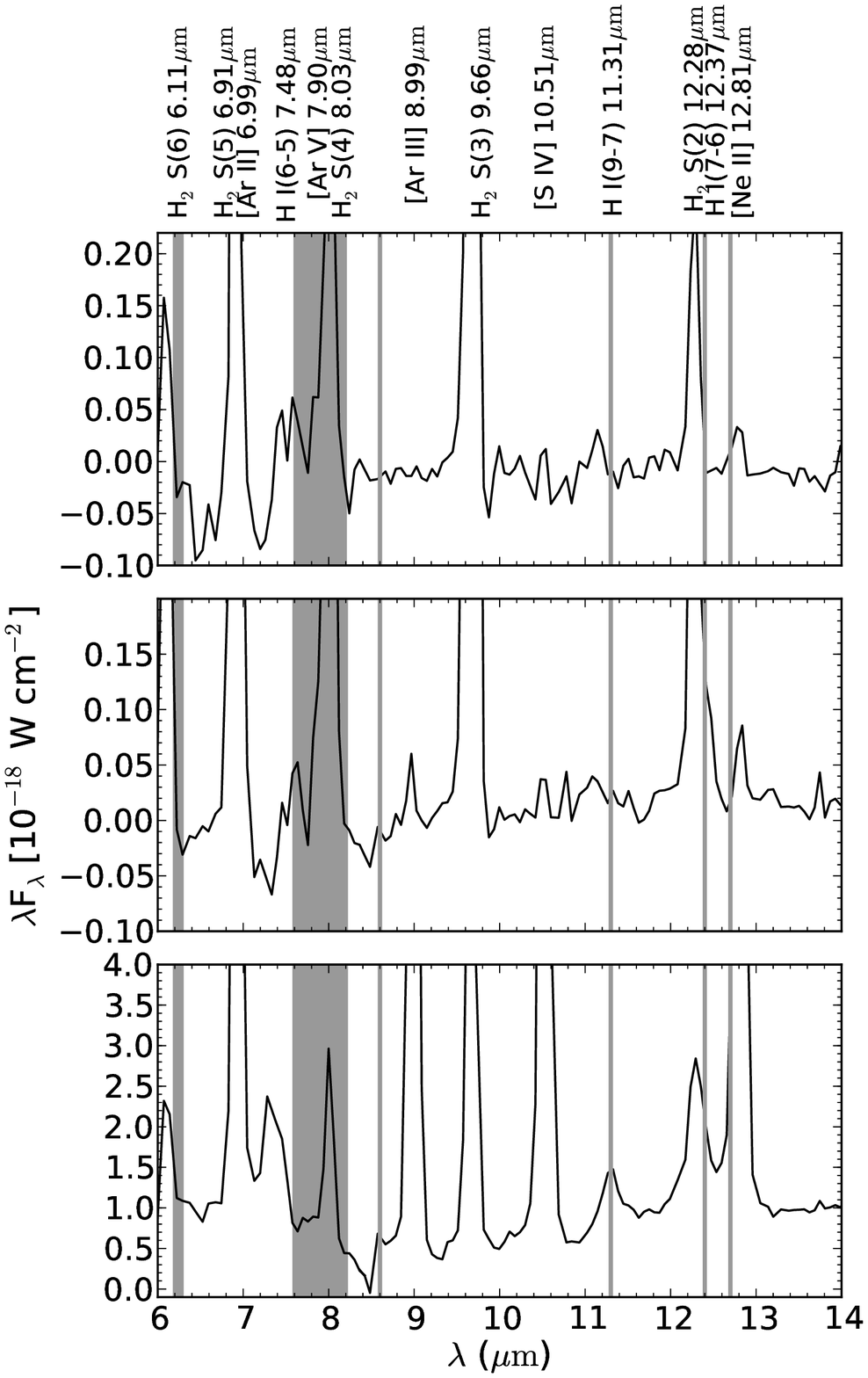}
\vspace{-0.5cm}
\caption{{\it Spitzer}/IRS spectra for three different positions in NGC~6720. Grey lines
and rectangles represent the usual peak intensity of the PAH features at 8.6, 11.3,
12.4, and 12.7 $\mu$m, and the range of variation for peak intensity of the features at 6.2
and 7.7 $\mu$m. \label{fig:dust_5}}
\end{figure}

The detection of PAHs is not clear in some nebulae, in particular those with spectra of low
spectral resolution, because of contamination with emission lines (see Fig.~\ref{fig:dust_5}
for some identifications). We see that in those PNe where PAHs are more conspicuous, such as
NGC~40 and Cn~1-5 (Fig.~\ref{fig:dust_3}), the broad feature at 11.3 $\mu$m is the PAH
feature most easily seen. Hence we used the presence of this broad feature as the main
criterion for the detection of PAHs.

Originally, PAHs were thought to appear only in those dust forming sources that are carbon-rich 
\citep{Whittet_03}, but in recent years these molecules have also been observed in oxygen-rich 
AGB and post-AGB stars, and in oxygen-rich PNe
\citep[see, e.g.,][]{Jura_06,Cerrigone_09,Guzman_11,Gielen_11}. 
These so-called mixed-chemistry PNe show PAHs and crystalline silicate features together 
in their infrared spectra \citep[see e.g.][]{Waters_98,Cohen_99}. We will come back 
to this issue in \S~\ref{orich}. 

Figure~\ref{fig:dust_5} shows the 6--14 $\mu$m IRS spectra of NGC~6720, which has observations
at three different locations. The upper panel shows the spectrum for a region that includes
part of the outer halo of the nebula; the middle panel shows the spectrum for an aperture
that crosses the halo; and the lower panel shows the spectrum for an aperture that crosses
the bright region close to the center of the nebula. We do not see evidence for PAHs, SiC, nor
amorphous silicates.

\subsubsection{O-rich features: amorphous and crystalline silicates}
\label{orich}

We observe the broad features associated with amorphous silicates in 3--5 
PNe of the 27 PNe with IRS/{\it Spitzer} spectra in the adequate wavelength range (see 
Figs.~\ref{fig:dust_3} and \ref{fig:dust_4}). 
DdDm~1, MyCn~18, and H~1-50 show the feature at 9.7 $\mu$m and some indication of the 
18 $\mu$m feature. We cannot rule out the presence of these features in IC~4846 and 
NGC~6210. Some PNe, like NGC~40, NGC~2392, and NGC~6210, show a small bump around 17--18
$\mu$m, but this feature is likely to be related with problems in the overlap of the
SH and LH spectra. Finally, \citet{BS_05} identified amorphous silicates in the {\it ISO}
spectra of NGC~6543 and, probably, in NGC~6153. 

We also identify the crystalline silicate features around 23$\mu$m, 27$\mu$m, and 33 $\mu$m 
in the IRS spectra of 12 PNe from the sample. We cannot rule out the presence of crystalline
silicates in IC~4846 and NGC~6818 due to the low resolution or poor signal-to-noise ratio of
their spectra. 
\citet{BS_05} used {\it ISO} spectra and detected crystalline silicates in NGC~6543, whereas
they could not rule out their presence in NGC~6153. Therefore, 13--16 nebulae from
the 21 PNe with available spectra in the required wavelength range show crystalline silicates.
The three PNe with amorphous silicates also show crystalline silicates. 

Six out of the thirteen PNe that show silicate features belong to the mixed-chemistry class,
objects that show both silicates and PAHs. Several scenarios have been proposed to explain
the simultaneous presence of silicates and PAHs in PNe \citep[see, e.g.,][]{PC_09}. 
One of the preferred explanations is that the silicates were formed and ejected 
before the star experienced the third dredge-up, when it was still O-rich, whereas PAHs
form later \citep{Waters_98}. According to this scenario, silicates are 
located in a O-rich disk/torus while PAHs occupy a more recent C-rich outflow. 
This could explain the mixed chemistry of Cn~1-5, where me measure C/O~$>1$. 
However, this explanation is not valid for stars that remain O-rich during their whole 
evolution, either because the third dredge-up does not occur due to their low initial 
masses or because the effect of the third dredge-up is counteracted by the hot bottom 
burning process. In these PNe, it might apply the explanation proposed by \citet{Guzman_11},
in which the PAHs form in an O-rich and dense torus after the CO molecules are dissociated. 

\section{Discussion}

The values of $12+\log(\mbox{O}/\mbox{H})$ in our sample of PNe go from 8.02 (for the halo 
PN DdDm~1) to 8.88 (for NGC~6620). 
If we consider the values of O/H found for the PNe that have different dust features, they 
cover the ranges: 8.26--8.87 (PAHs), 8.26--8.68 (SiC or 30 $\mu$m feature),
8.02--8.87 (silicates), and 8.45--8.87 (mixed-chemistry objects: silicates and PAHs). Thus
the sample PNe with silicate dust grains seem to arise from a broad population that includes
the halo PN, DdDm~1, PNe from the Galactic disk, and objects that might belong to the bulge,
like H~1-50 and M~2-31 \citep{Wang_07}. Our PNe with SiC and the 30
$\mu$m feature seem  to arise from a more homogeneous population in the Galactic disk; only one
of them, M~1-20, might belong to the bulge. This agrees with the results of
\citet{Stanghellini_12}, who found that six of their compact Galactic PNe with C-rich dust
features, for which they could estimate the dust temperature, follow a well defined sequence in
their plots of dust temperature versus infrared luminosity or physical radii.
\citet{Stanghellini_12} conclude that this is an evolutionary sequence and that the
progenitors of these PNe covered a narrow range in initial mass and metallicity. The high oxygen
abundances found for the mixed-chemistry PNe also agree with the results of
\citet{Stanghellini_12}, since they found that these nebulae are concentrated towards the
Galactic center and are absent from the Magellanic Clouds.

\begin{figure*}
\begin{center}
\includegraphics[width=16cm,trim = 40 60 20 50,clip =yes]{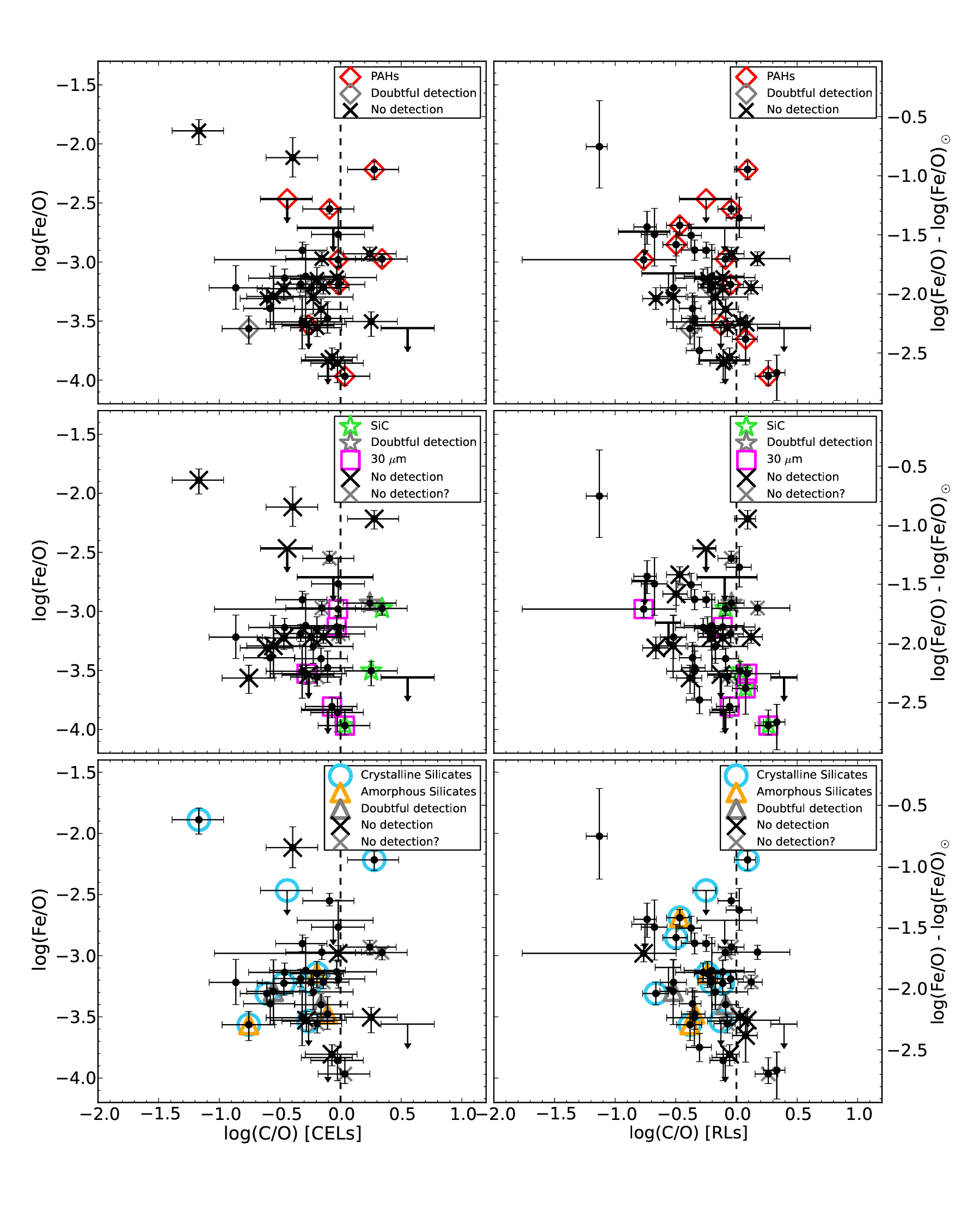}
\caption{Values of Fe/O (left axis) and the depletion factors for Fe/O 
($[\mbox{Fe/O}]= \log(\mbox{Fe/O}) - \log(\mbox{Fe/O})_{\odot}$, right axis) as a function of 
the C/O abundance ratios obtained from CELs (right panels) and RLs (left panels) for all 
the PNe with available data. We use the symbols identified in each panel to mark those PNe
with detections and non-detections of different infrared dust features. \\
(A color version of this figure is available in the online journal.)\label{fig:disc_1}}
\end{center}
\end{figure*}

The range of iron depletions we find in our sample of PNe covers about two orders of
magnitude, suggesting differences in the formation and evolution of dust grains from one PN
to another. 
As in \citet{DelgadoInglada_09}, we explored possible correlations between the 
iron depletions and different nebular and stellar parameters such as the electron density, 
the nebular radius, the surface brightness, the effective temperature of the central star, or the 
nebular morphology. In agreement with \citet{DelgadoInglada_09}, we do not find any obvious
 correlation.
We investigate here if this large range of iron depletions is related to the type of dust
grains found in the PN (i.e., C-rich or O-rich dust grains) or to the dominant chemistry
present in the ionized gas (i.e., if the computed C/O ratio is lower or greater than one).  
In Figure~\ref{fig:disc_1} we display the values of Fe/O and the depletion factors for Fe/O
as a function of the C/O abundance ratios, derived from CELs and RLs, for all the PNe in our
sample with available data. Note that the sample PNe can have two, one, or no values for C/O
(depending on whether the required CELs and RLs are measured or not) and hence not all the PNe
in Table~\ref{tab:3} appear in Figure~\ref{fig:disc_1}, and some objects only appear in the
panels at the right (left).
We present the results for the values of Fe/O derived with equations~(\ref{eq:2}) and
(\ref{eq:3}), but the values of Fe/O implied by equation~(\ref{eq:1}) lead to similar results.

We also identify in Figure~\ref{fig:disc_1} the PNe showing dust features in their infrared
spectra, as well as those with doubtful or no detection of dust features. Note that, given the
uncertainties related to the calculation of C/O from emission lines, the type of dust grains
(O-rich or C-rich excluding PAHs) might provide better information on whether the intrinsic
value of C/O in the nebular gas is above 1 or not
\citep[unless grain components like SiC can form in slightly O-rich conditions, see][]{Bond_10}.

The upper panels of Figure~\ref{fig:disc_1} show the PAH identifications. The first studies on 
the relation between PAHs and the C/O abundance ratios found a trend of increasing 
strength of these features as C/O increases \citep[][and references therein]{Cohen_05},
and it was suggested that these molecules are only present in PNe with C/O~$>1$. 
Here we find some PNe with intense PAH features and low C/O ratios, such as NGC~7026 
or NGC~6439. Besides, PAH emission is not restricted to PNe with C/O~$>1$, in agreement 
with the idea that PAH molecules may form both in C-rich and O-rich 
environments \citep{Guzman_11}. We also see from this figure that PNe with PAHs are distributed 
over the whole range of iron depletions, suggesting that the presence of these molecules has
no direct relation with the highest and lowest iron depletions found in the ionized gas. 

The middle panels of Figure~\ref{fig:disc_1} identify those PNe with SiC or the broad
feature at 30~$\mu$m, usually associated with dust grains formed in a C-rich environment.
Most of these PNe are located close to the region where C/O~$>1$.
NGC~40 has a C/O value from RLs that is not consistent within the
errors with  C/O~$>1$, but its C/O value from CELs is consistent with C/O = 1.
On the other hand, \citet{Bond_10} find in their simulations of terrestrial planet
formation that solid SiC can form when the C/O ratio is above 0.8. In this case, the range
of C/O ratios for nebulae with C-rich dust can be somewhat wider.

The lower panels of Figure~\ref{fig:disc_1} show the PNe with amorphous or crystalline
silicates in their spectra. These dust features seem to be very common in our sample. All but
one of those PNe with silicate features have values of C/O that are compatible with an O-rich
environment. Cn~1-5 is the exception, showing silicates with a C/O value clearly above one,
regardless of what lines are used in the abundance determination and regardless of the 
adopted ICF. One possible explanation
for this is that the C$^{++}$/O$^{++}$ values derived from CELs and RLs are seriously wrong
for this nebula. Another option, already mentioned in \S~\ref{sec:dust}, is that the computed
C/O value in the ionized gas is correct, and the prominent silicate features are revealing an
O-rich past where the silicate grains formed. 

If we consider as C-rich those PNe with SiC or the 30~$\mu$m feature and as O-rich those with
silicates, we can see in Figure~\ref{fig:disc_1} that both C-rich and O-rich PNe cover a wide
range of iron depletions. However, C-rich nebulae are not present in the region of lowest iron
depletions (highest gaseous iron abundances), whereas O-rich PNe do not appear at the highest
depletions. The PNe with the highest iron depletions that have identifications of dust features
are IC~418 and NGC~3242, two PNe with C-rich dust features and no silicates. On
the other hand, the PNe with the lowest depletions that also have identifications of dust
features are Cn~1-5, DdDm~1, M~1-42, and NGC~2392, four PNe that show silicates and no C-rich
dust features. In order to check whether this finding is due to small-number statistics, we
performed a Kolmogorov-Smirnov (KS) test that compares the distributions of iron depletions for
C-rich and O-rich PNe (as defined by the presence of SiC or the 30~$\mu$m feature, versus the
presence of silicates). Using the values of Fe/O in Figure~\ref{fig:disc_1} we find that the
p-value (the probability of finding different distributions of Fe/O for C-rich and O-rich PNe
if both arise from the same distribution) is 4\%. We also performed the KS test using the
values of Fe/O implied by equation~(\ref{eq:1}) and the values of Fe/H implied by both
ionization correction schemes, finding probabilities in the range 0.2\% to 7\%, with the
values of Fe/H implying the lowest values (0.2\% and 1\%). Therefore, the difference is
significant at the 0.2--7\% level.

An independent way to test this relation of iron depletions with C/O is by studying the
correlation between the iron abundances and the values derived for C/O. We studied this
correlation considering the two values of Fe/H and Fe/O implied by the two different
ionization correction schemes we use, and the values of C/O implied by either CELs or RLs
(making a total of 8 cases). We find weak Spearman rank correlation coefficients, in
the range $-0.4$ to $-0.1$, and p-values that go from 0.1\% (for iron abundances derived
from eq.~(\ref{eq:2}) and (\ref{eq:3}) and C/O values implied by RLs) to 30\%. Hence,  the
relation between iron depletions and C/O abundances, although weak, seems to be
significant. This relation could arise from changes in the composition of dust grains for
environments characterized by different values of C/O.

Our results do not agree with the model results of \citet{Ferrarotti_06} for dust
production in AGB stars. Whereas most of our objects show high iron depletion factors,
their calculations only show large degrees of condensation for iron, above 80\%, in
objects where the value of C/O is extremely close to 1 (see their Figs.~A.3 and A.4).
As discussed by \citet{Mauron_10}, who find strong depletions of elements like Fe and Ca in
the circumstellar envelope of an AGB carbon star, adsorption of metals at the
surface of the dust grains may be as important as the initial dust condensation processes
considered by the models of dust formation in stellar envelopes.

\section{Summary and conclusions}

We have studied a sample of 56 PNe, belonging to the Galactic bulge, the halo, and the disk.
The nebulae have available optical spectra of good quality and we constrain their iron
abundances using [\ion{Fe}{3}] lines and the ionization correction scheme derived by
\citet{Rodriguez_05}. The iron depletion factors we find imply that most of the studied PNe
have less than 10\% of their iron atoms in the gas, the missing iron atoms are presumably
condensed into dust grains.  

We have calculated the C/O abundance ratios using both optical RLs (for 49 PNe) and
ultraviolet and optical CELs (for 39 PNe). The differences between the two estimates
reach $\sim0.7$ dex in some objects.
Most of the sample PNe show C/O~$<1$ which, according to theoretical models, indicates that 
they could either descend from stars with masses below $\sim1.5$ M$_{\odot}$ or from stars 
with masses above 4--5 M$_{\odot}$. Those PNe with C/O~$>1$ are probably the result of 
intermediate mass stars.

We have also studied the infrared spectra of the PNe with available data in order to identify
the following features: PAHs, SiC, the broad band at 30 $\mu$m, and amorphous and crystalline
silicates. We find PAHs in 12--14 of the 33 PNe with infrared spectra. The presence of these
features is not restricted to PNe with C/O~$>1$. Silicates are found in 13--16 PNe; SiC and/or
the 30 $\mu$m feature, which is also associated with C-rich environments, in 7--9 objects. The
presence of silicates and SiC or the 30 $\mu$m feature does not agree in all cases with the
values of C/O derived from emission lines, maybe reflecting the uncertainties related to the
determination of the C/O abundance ratio. Note also that the C/O value we measure can
be affected by the depletion of C or O atoms into dust grains.

The oxygen abundances, $12+\log(\mbox{O}/\mbox{H})$, found for the PNe that have different dust
features cover the ranges: 8.26--8.87 (PAHs), 8.26--8.68 (SiC or 30 $\mu$m feature), 8.02--8.87
(silicates), and 8.45--8.87 (mixed-chemistry objects: silicates and PAHs). Therefore, the
sample PNe with silicate dust grains seem to arise from a broad population that includes our
halo PN, PNe from the Galactic disk, and objects that might belong to the bulge. Our PNe with
SiC and the 30 $\mu$m feature seem to arise from a more homogeneous population in the Galactic
disk, while the mixed-chemistry objects arise from PNe of relatively high metallicities,
results that agree with those found by \citet{Stanghellini_12} using measurements of the dust
temperature and the distribution in the Galaxy of their sample objects.

We find a relation between the iron depletions and the type of chemistry in the nebulae, 
C-rich or O-rich. Both C-rich and O-rich PNe cover a wide range of iron depletions, but the
PNe with the highest iron depletions have C-rich dust features (SiC and/or the 30 $\mu$m
band) whereas those PNe with the lowest iron depletions have silicates in their infrared
spectra. In accordance with this result, we find a weak but significant
anticorrelation between the iron abundances and the C/O abundance ratios derived from
emission lines. Some kind of correlation is expected from the different molecules and dust
compounds that should form in environments with different values of C/O.

\begin{acknowledgements}
We thank the anonymous referee for valuable comments that helped to improved the paper.
We acknowledge support from Mexican CONACYT grants 50359-F and 131610-F. We have used NASA's
Astrophysics Data System, and the SIMBAD database operated at CDS, Strasbourg, France. This
work is based in part on archival data obtained with the {\it Spitzer}  Space Telescope, which
is operated by the Jet Propulsion Laboratory, California Institute of Technology under a
contract with NASA.
\end{acknowledgements}

\bibliographystyle{apj}

\end{document}